\documentclass[reprint,superscriptaddress,amsmath,amssymb,aps,prb]{revtex4-1}

\usepackage{natbib}
\usepackage{graphicx}
\usepackage{dcolumn}
\usepackage{bm}
\usepackage{gensymb}
\usepackage{textcomp}
\usepackage{xfrac}
\usepackage{amsmath}
\usepackage{bbold}
\usepackage{xcolor}
\usepackage{placeins}

\begin{document}
\title{Doped NiO: the Mottness of a charge transfer insulator}
\thanks{Notice: This manuscript has been authored by UT-Battelle, LLC, under Contract No. DE-AC0500OR22725 with the U.S. Department of Energy. The United States Government retains and the publisher, by accepting the article for publication, acknowledges that the United States Government retains a non-exclusive, paid-up, irrevocable, world-wide license to publish or reproduce the published form of this manuscript, or allow others to do so, for the United States Government purposes. The Department of Energy will provide public access to these results of federally sponsored research in accordance with the DOE Public Access Plan (\url{http://energy.gov/downloads/doe-public-access-plan}).}

\author{Friederike Wrobel}
\affiliation{Materials Science Division, Argonne National Laboratory, Lemont, Illinois 60439, USA}

\author{Hyowon Park}
\affiliation{Department of Physics, University of Illinois at Chicago, Chicago, IL 60607, USA}
\affiliation{Materials Science Division, Argonne National Laboratory, Lemont, Illinois 60439, USA}

\author{Changhee Sohn}
\affiliation{Center for Nanophase Materials Sciences, Oak Ridge National Laboratory, Oak Ridge, Tennessee 37831, USA}
\affiliation{Department of Physics, Ulsan National Institute of Science and Technology, Ulsan, Korea}

\author{Haw-Wen Hsiao}
\affiliation{Department of Materials Science and Engineering, University of Illinois at Urbana-Champaign, Urbana, IL 61801, USA}

\author{Jian-Min Zuo}
\affiliation{Department of Materials Science and Engineering, University of Illinois at Urbana-Champaign, Urbana, IL 61801, USA}

\author{Hyeondeok Shin}%
\affiliation{Computational Science Division, Argonne National Laboratory, Argonne, Illinois 60439, USA}


\author{Ho Nyung Lee}
\affiliation{Materials Science and Technology Division, Oak Ridge National Laboratory, Oak Ridge, TN 37831, USA}
\affiliation{Center for Nanophase Materials Sciences, Oak Ridge National Laboratory, Oak Ridge, Tennessee 37831, USA}

\author{P. Ganesh}
\affiliation{Center for Nanophase Materials Sciences, Oak Ridge National Laboratory, Oak Ridge, Tennessee 37831, USA}


\author{Anouar Benali}
\affiliation{Computational Science Division, Argonne National Laboratory, Argonne, Illinois 60439, USA}
\affiliation{Leadership Computing Facility, Argonne National Laboratory, Argonne, Illinois 60439, USA}

\author{Paul R.C. Kent}
\affiliation{Center for Nanophase Materials Sciences, Oak Ridge National Laboratory, Oak Ridge, Tennessee 37831, USA}
\affiliation{Computational Sciences and Engineering Division, Oak Ridge National Laboratory, Oak Ridge, Tennessee 37831, USA}

\author{Olle Heinonen}
\email{heinonen@anl.gov}
\affiliation{Materials Science Division, Argonne National Laboratory, Lemont, Illinois 60439, USA}
\affiliation{Center for Hierarchical Material Design, Northwestern-Argonne Institute for Science and Engineering,Northwestern University, Evanston, Illinois 60208, USA}

\author{Anand Bhattacharya}
\affiliation{Materials Science Division, Argonne National Laboratory, Lemont, Illinois 60439, USA}
%




\date{\today}

\begin{abstract}
The evolution of the electronic structures of strongly correlated insulators with doping has long been a central fundamental question in condensed matter physics; it is also of great practical relevance for applications. We have studied the evolution of NiO under hole {\em and} electron doping at low doping levels such that the system remains insulating using high-quality thin film and a wide range of experimental and theoretical methods. The evolution is in both cases very smooth with dopant concentration. The band gap is asymmetric under electron and hole doping, consistent with a charge-transfer insulator picture, and is reduced faster under hole than electron doping. For both electron and hole doping, occupied states are introduced at the top of the valence band. The formation of deep donor levels under electron doping and the inability to pin otherwise empty states near the conduction band edge is indicative that local electron addition and removal energies are dominated by a Mott-like Hubbard $U$-interaction even though the global bandgap is predominantly a charge-transfer type gap.
\end{abstract}


\maketitle


\section{Introduction}
\label{sec:Intro}

Predicting the behavior of materials with strong electronic correlations is a long-standing problem in condensed matter physics and materials sciences. Strong electronic correlations give rise to a host of interesting properties, {\em e.g.,} colossal magnetoresistance and high-T$_c$ superconductivity. 
In systems with partially filled bands, electronic interactions can lead to insulating states. For example, at half fillings arbitrarily weak interactions can lead to Slater-type antiferromagnets\cite{slater1951} or spin density waves\cite{overhauser1962}. Long ago, de Boer and Verwey\cite{de1937} pointed out that some materials with incompletely filled 3$d$ bands, such as NiO, are unexpectedly insulating. This prompted Peirels and Mott\cite{mott1937} to discuss the role of electron interactions in driving materials that should be metallic by simple electron counting to an insulating state. It is now well known that strong electronic correlations can give rise to an antiferromagnetic insulating state when simple electron counting and band theory suggest that the material should be a metal. This is, for example, typically the case for the normal parent states of high-T$_c$ superconductors\cite{lee2006}.
These insulators are typically referred to as Mott insulators. Some of the simplest ones are transition metal oxides (TMOs), such as NiO and VO$_2$. In these, the relatively localized $3d$ electrons are the source of the strong correlations. TMOs often exhibit metal-to-insulator transitions (MITs) as a function of doping, strain, or temperature; this functionality has made TMOs attractive for applications in electronics and information technologies that exploit the MIT, so-called Mottronics~\cite{inoue2008}.

A central question in most potential applications, as well as from a fundamental physics perspective, is understanding how the transition from insulator to a metallic state occurs under doping. As is well known, in weakly interacting semiconductors such as Si and Ge, dopants introduce states in the band gap, either donor states close to the conduction band edge, or acceptor states close to the valence band edge. These states are well described as hydrogen-like but with an effective mass given by the curvature of the electronic bands rather than the electron mass. As the dopant concentration increases, these hydrogen-like states start to overlap, forming bands. A key point here is that the electron bands of the semiconductor remain unchanged under doping, a so-called rigid band model -- all the dopants do is to add states in the energy gap. 

\begin{figure*}
    \includegraphics[width=17.9cm]{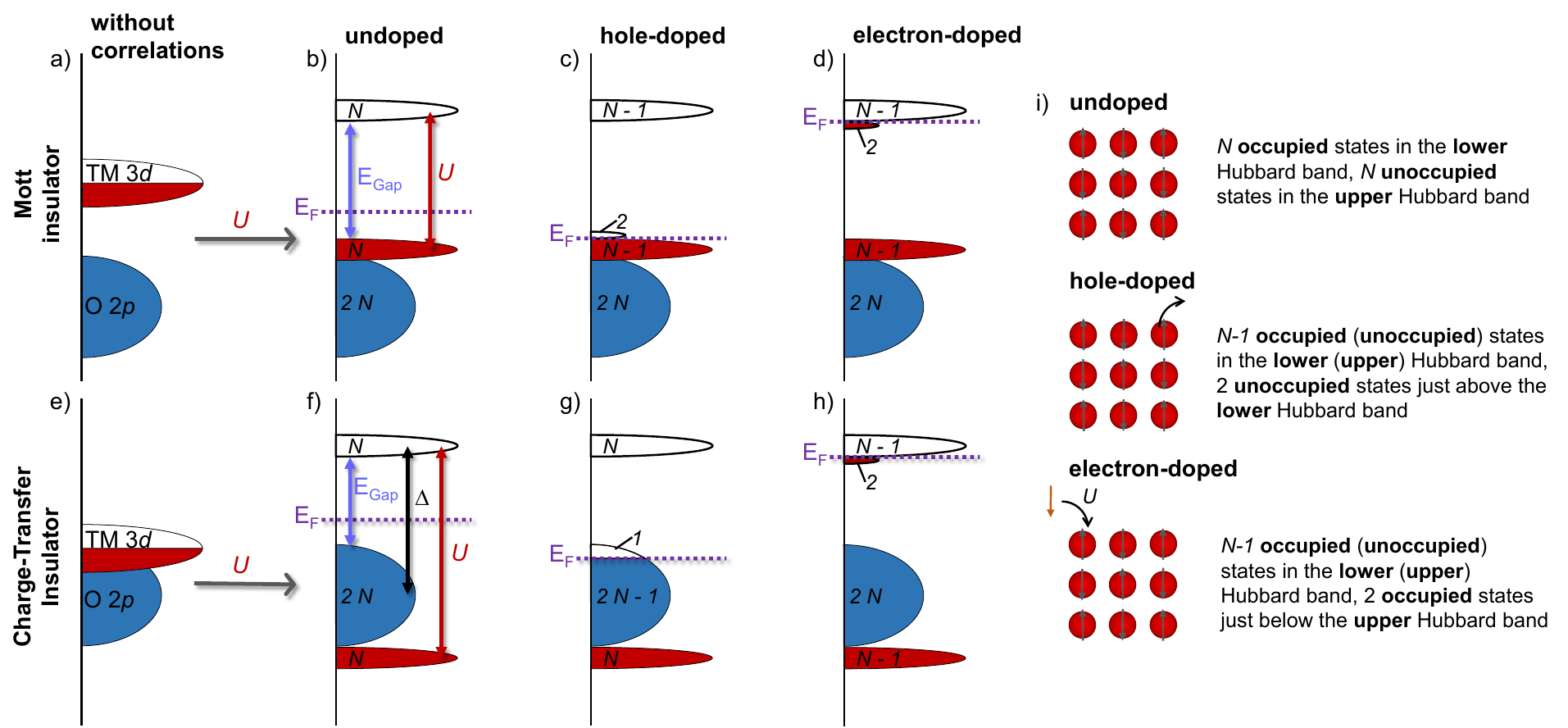}
    \centering
    \caption{\label{fig:MottCT} Cartoon of electron densities of states (DOS) of an archetypical Mott insulator (upper panels, a-d) and charge-transfer insulator (lower panels, e-h). a) and e) show the hypothetical DOS without correlations, where TM denotes a transition metal. 
    On-site correlations, i.e., the Hubbard repulsion $U$, lead to a splitting of the 3$d$ band into an upper and lower Hubbard band, each occupied with $N$ electrons (b and f). There are 2$N$ states in the O 2$p$ band, which is fully occupied. E$_{\mathrm{F}}$ denotes the Fermi level, E$_{\mathrm{Gap}}$ the band gap, $\Delta$ the charge transfer energy, i.e., the energy needed to transfer an electron from the O 2$p$ band to the upper Hubbard band.
    c) Removal of one electron from a Mott insulator, i.e., hole-doping, creates one TM site with two unoccupied states just above the Fermi level and removes one state each from the upper and lower Hubbard band.
    g) Removal of one electron from a CT-insulator creates a hole in the O 2$p$ band but leaves the upper and lower Hubbard band untouched.
    d) and h) Addition of one electron to a Mott Hubbard or CT insulator, i.e., electron-doping, creates one TM site with two occupied states just below the Fermi level and removes one state each from the upper and lower Hubbard band.
    i) Local picture of hole- and electron-doping of a Mott insulator.}
\end{figure*}

In contrast, the behavior of correlated oxides under doping is much more complicated. First of all, the standard theoretical work horse for electronic structure, density functional theory (DFT), generally fails to accurately capture the effect of electronic correlations. 
Extensions of DFT to better include local correlations, such as DFT+U in which a Hubbard $U$-parameter is added to mimic on-site Coulomb repulsions~\cite{anisimov1991}, generally open up a gap (as visualized by the sketch in Fig.~\ref{fig:MottCT} b and f) and improve on ground state properties. However, results for properties involving excitations and charge doping, such as photoemission spectra, still tend to be in poor agreement with experiments. Other, more expensive techniques better suited to address strong correlations, in particular dynamic correlations, such as dynamical mean field theory (DMFT) or quantum Monte Carlo (QMC) methods are needed. In general, as dopants are added, the entire electronic structure is modified, especially conduction and valence bands.

Fig.~\ref{fig:MottCT} shows the density of states (DOS) of an archetypical Mott insulator (a-d) and charge-transfer insulator (e-h) upon hole- (c and g) and electron-doping (d and h). Due to strong on-site Coulomb interactions, adding a second electron to a transition metal (TM) 3$d$ level, comes at a cost $U$, which opens up a gap between unoccupied $d$ states (upper Hubbard band) and occupied $d$ states (lower Hubbard band) as depicted in Fig.~\ref{fig:MottCT}b and f. Upon hole- and electron-doping, the number of states in the upper Hubbard band changes together with the number of states in the lower Hubbard. In addition, new states may appear in the gap. This is usually seen experimentally as a shift in spectral weights in, for example,  optical conductivity. In addition to the weight shift depicted in Fig.~\ref{fig:MottCT}, the energy separation between the different bands may shift as well, which is observed experimentally as a peak shift in optical conductivity. Both, the transfer of weight and the shift of energy levels are unique to Mott and charge-transfer insulators as opposed to classical semiconductors.

The central aim of our work is to elucidate the evolution of the electronic structure upon doping of NiO, which we use as a prototypical correlated oxide, using a broad range of experimental and theoretical tools. We note that our interest is on low dopant concentrations such that the system remains insulating and the band gap has not collapsed.

Nickel oxide, NiO, is perhaps the simplest correlated TMO. It has a rock salt structure with a lattice constant of about 4.18\,{\AA}. Pure NiO has a N\'eel temperature $T_N=523$\,K and an optical band gap of about 3.5 - 3.7\,eV measured by optical reflectivity~\cite{newman1959,powell1970} or optical  absorption~\cite{sasi2002,zhang2006}; the band gap remains at temperatures far above the N\'eel temperature. 
The fundamental behavior of pure NiO falls generally in the {\em Intermediate region} in the Zaanen-Sawatzky-Allen theory of transition-metal compounds\cite{zaanen1985}.  
NiO can be hole-doped, for example, through Ni deficiency, or through substitutional doping with Li or K. While there is not much work on electron-doped NiO in the literature~\cite{reinert1995,oka2012,kerli2014investigation}, there is a relative wealth of work on hole-doped NiO.   

Feinleib and Adler~\cite{feinleib1968} argued that conduction in Li-doped NiO occurred primarily in the oxygen $2p$ band, while Ni $3d$ states are localized and the Ni $4s$ states are unoccupied at high energies and therefore only contribute to conduction at high temperatures. In their scenario the optical absorption edge at about 4\,eV is caused by $2p\to4s$ transitions, with optical absorption below 4\,eV mainly due to $d-d$ transitions on a single nickel ion.

Kuiper, {\em et al.}\cite{kuiper1989character} used x-ray absorption spectroscopy to study NiO and Li-doped NiO, with Li doping in the range of 5\% to 50\%. Based on the emergence in the oxygen K-edge spectroscopy of an oxygen 1$s$ absorption peak at about 528.5~eV with doping, they concluded that holes are primarily located on oxygen 2$p$ bound to Li$^+$ dopants, forming states in the gap close to the valence band edge. This picture was confirmed by H\"ufner {em et al.}\cite{hufner1992acceptors} using photoemission and inverse photoemission spectroscopy on 2\% Li-doped NiO.

van Elp, Eskes, Kuiper, and Sawatzky~\cite{van1992} used x-ray photoemission spectroscopy (XPS) and bremsstrahlung isochromat spectroscopy (BIS) to study Li-doped NiO and compared their results with cluster calculations. They concluded that the doped hole is of mainly oxygen character. The XPS valence band spectra showed mainly a reduction of intensity but no additional peaks upon relatively high doping (40\%). 
The BIS (conduction band) spectra showed more dramatic evolution. They find a $d^9$ conduction band peak, i.e., the upper Hubbard band, at 5\,eV. With Li-doping, impurity states arise below the $d^9$-peak and their spectral weight grows linearly with doping while the spectral weight of the $d^9$-peak decreases simultaneously.

Kunes, Anisimov, Lukoyanov, and Vollhardt~\cite{kunevs2007prb} used DMFT to study a down-folded eight-band Hamiltonian of pure NiO and heavily hole-doped (38\%) NiO, and calculated spectral densities resolved on O $2p$ and Ni $3d$ orbitals.  
They found valence band peaks due to Ni $3d$ closest to the valence band edge, and a smaller O $2p$ peak close to the band edge, and below the Ni $3d$ peak. 
The conduction band peak was dominated by Ni $3d$ ($e_g$ character). Upon doping, the $e_g$ spectral function changed significantly: the Mott gap is filled, but with Hubbard band peaks preserved as distinct features. Further doping transfers spectral weights from the upper and lower Hubbard bands to an $e_g$ quasiparticle peak.

Kunes, Anisomov, Skornyakov, Lukoyanov, and Vollhardt~\cite{kunevs2007prl} compared calculated DMFT  spectral densities $A(\mathbf{k},\omega)$ for pure NiO 
with angle-resolved photoemission spectroscopy (ARPES) data for NiO. Along the $\Gamma-X$ and $\Gamma-K$ lines in the 1st Brillouin zone, they found relatively flat bands at $-2$\,eV and $-4$\,eV below the valence band edge. 
They concluded that $p-d$ hybridization leads to mixed-character bands at $-4$\,eV and $-2$\,eV with a broadening of the lower Hubbard band,  
and that the $p$-bands acquire finite widths due to coupling to $d$-band.

Chen and Harding\cite{chen2012} used DFT+U and the HSE06 hybrid exchange-correlation functional\cite{heyd2005} to calculate DOS and charge densities for Li-doped NiO at 12.5\% doping. The found three defects states in the gap. Notably the defect states were pushed from the valence band edge below the Fermi energy as the oxygen octahedra around the Li dopant relaxed in a Jahn-Teller like distortion. The defect states had mostly hybridized O 2$p$ and Ni 3$d$ character, with the charge density resided on Ni$^{3+}$ next-nearest neighbor to the dopant.

Finally, Shinohara {\em et al.}~\cite{shinohara2015} 
use reduced density matrix theory to study the insulator-to-metal phase transition in NiO under hole doping. They concluded that within this approach, a redistribution of spectral weights drives the transition; Ni $e_g$ states spectral weights drive towards the chemical potential, while the Ni $3d$ $t_{2g}$ spectral weight remains almost unchanged.

Reinert {\em et al.}~\cite{reinert1995} investigated both hole- and electron-doped thin films obtained by doping with Li and by depleting NiO thin films of oxygen. They found that electron-doping gives rise to defect states 0.6\,eV and 1.7\,eV above the valence band edge, and argued that these were both Ni$^0$ states, with the lower one an occupied pinning center. Oka {\em et al.}~\cite{oka2012} used DFT+U to calculate the band structure of hole- (electron-) doped NiO obtained by placing a Ni (O) vacancy in a Ni$_{32}$O$_{32}$ supercell. In qualitative agreement with Reinert {\em et al.}~\cite{reinert1995}, they found that the O vacancy gives rise to an occupied state about 0.8\,eV above the valence band edge, in addition to another (empty) state about 1.5\,eV higher in energy but about 1\,eV below the conduction band edge.

Kerli, Alver, and Yaykasli\cite{kerli2014investigation} measured the optical conductivity of thick (about 1~$\mu$m) In-doped NiO films grown using airbrush spraying, with dopant concentrations from 0\% to 10\%. They found an {\em increase} in the optical band gap from 3.65~eV at 0\% doping to 3.99~eV at 8\%. The increased band gap was attributed to decreased relative oxygen content with In$^{3+}$ occupying Ni vacancies. In addition, they observed a lattice contraction upon doping, which is not consistent with In having a larger ionic radius than Ni. It is therefore likely that the effects they observed were related to sample-dependent extrinsic properties because of the relatively poor sample quality rather than intrinsic properties.

The summary, for our purposes here, is then as follows for low dopant concentrations so that the gap has not collapsed. (i) in pure NiO, the states at the valence band edge are mostly O 2$p$ hybridized with Ni 3$d$ states while the states at the conduction band edge are mainly Ni 3$d$; (ii) hole-doping with Li gives rise to mid-gap defect states, the lowest one about 1~eV above the valence band edge. The hole states are mixed O 2$p$ and Ni 3$d$ character and are localized on Ni. (iii) Electron doping through oxygen depletion gives rise to midgap states, probably Ni$^0$, with the lowest one acting as a pinning center. (iv) The electronic structure upon doping with substitutional impurities, such as In, remains unclear.

We have grown high-quality K-doped (hole-doping) and In-doped (electron-doping) NiO thin films, using molecular beam epitaxy, with doping ranges up to about 10\%. These are doping ranges low enough that the doped system remains insulating. We have studied the band gap evolution upon doping using spectroscopic ellipsometry, and compared our results using DFT+U and DMFT. Key results in our work are first of all that the doping introduces a spectral peak just above the valence band edge, and spectral weight is transferred from transitions from a mainly O $2p$ peak to the upper Hubbard band (Ni $3d$ $e_g$ character) to this new peak. The evolution of the band gap with doping is asymmetric in hole and electron doping, with a much more rapid decrease of the band gap for hole doping than for electron doping, confirming the charge-transfer nature of NiO. We find that both hole and electron doping lead to occupied states in the gap, both at the valence band edge. L\"owdin charge analysis of the Ni atoms based on DFT+U calculations show that 
electron doping leads to an increase in the occupancy of Ni $d$ orbitals, with a concomitant reduction in spin polarization in the Ni shell nearest the dopant. Hole doping, on the other hand, leads to a slight reduction in the Ni $d$-orbital occupancy. 

\section{Methods}
\label{sec:Methods}

\subsection{Sample preparation and characterization}
\label{subsec:Prep}

To grow potassium (indium) doped NiO thin films, nickel and potassium (indium) were evaporated from effusion cells and deposited on the atomically flat (100) surface of rocksalt-type MgO substrates with ozone-assisted molecular beam epitaxy. Before growth, the substrates were cleaned with acetone and isopropanol in an ultrasonic bath and annealed for few minutes at 600\,\celsius\ in a $1\times 10^{-6}$\,mbar ozone atmosphere. The samples were grown at a temperature of 200\,\celsius\ and in a $4\times 10^{-7}$\,mbar ozone atmosphere.

A high-resolution, four-circle diffractometer (Bruker D8 Discover equipped with a point detector) was used to measure x-ray diffraction with Cu $K_{\alpha 1}$ radiation.
Film thickness and $c$ axis parameter were extracted from low-angle x-ray reflectivity and $\theta -2\theta $ scans through the (002) reflections of film and substrate, respectively.

The composition of all samples was measured with Rutherford backscattering (RBS) and analyzed with SIMNRA software \cite{mayer1997}. Exemplary spectra are shown in the Appendix.

Two samples, one undoped NiO thin film and one doped with 3.6\%~potassium, were prepared for transmission electron microscopy (TEM) by focused ion beam (FIB) cross-section and studied by scanning TEM combined with electron energy loss spectroscopy (STEM-EELS) using an aberration-corrected scanning transmission electron microscope (Themis Z, Thermo-Fisher Scientific), operated at 80 kV. The EELS spectra were collected using dual-detector, one for the low-loss region including the zero-loss peak, and the other for oxygen $K$-edge or nickel $L_{2,3}$-edges. Additionally, we used the electron monochromator to reduce the zero-loss peak width to 0.2-0.3\,eV in FWHM. During EELS acquisition, the electron probe was scanned over the targeted area with a step size of 1\,nm (the electron probe size is much less than 1\,nm). EELS spectra were recorded at each probe position. The EELS collection angle was 28\,mrad, which is slightly larger than the beam convergence angle (25\,mrad). The spectra presented in Fig.~\ref{fig:NiEELS} and Fig.~\ref{fig:OEELS} were averaged over the scanned area larger than 400\,nm$^2$ with $t/\lambda$ of 1.0 (doped sample) or 0.2 (undoped sample); $t$ is the thickness of the sample and $\lambda$ is the mean free path of the inelastically scattered electrons. The estimated thickness of the specimens are 63\,nm for the doped sample and 15\,nm for the undoped sample based on the assumption that $\lambda$ is identical in both samples. In reality, the thicknesses for both samples should be similar which can be explained by differing values for $\lambda$ in the two samples. Therefore, we can give a lower bound of 15\,nm and an upper bound of 63\,nm. This corresponds to a minimum of $\sim 83\times 10^3$ unit cells and a maximum of $\sim 35 \times 10^4$ unit cells in the averaged area.

To check the sample homogeneity, we performed smaller area scans over areas of 16\,nm$^2$ (corresponding to $\sim 3 \times 10^3$ -- $15 \times 10^3$ unit cells), positioned near the top and the middle of the films. These scans show that the EELS spectra are nearly identical with a difference smaller than the experimental noise. A map of the Ni and O peak positions based on these smaller area scans is shown in the inset or the bottom panel of Fig.~\ref{fig:NiEELS} and Fig.~\ref{fig:OEELS}.

To measure the spectral energy shift, we calibrated the energy loss in each spectrum by using the recorded zero-loss peak as the energy reference, where the energy shift is applied to the core-loss spectrum, which was simultaneously recorded with the low-loss spectrum. After energy loss calibration, the nickel $L_3$ edge, for example, has a standard deviation of 0.05\,eV over the scanned area.

The ellipsometric angles $\Psi$ and $\Delta$ were measured with a UVISEL Spectroscopic ellipsometer from Horiba Jobin Yvon in the range of 0.6–-6.5\,eV and at an angle of 70\degree. The refractive index $n$ and extinction coefficient $k$ were obtained using the built-in software, assuming an isotropic, homogeneous film with the thickness measured by x-ray reflectivity.

\subsection{DFT+DMFT}
\label{subsec:DFT+DMFT}

Geometries for K(In)-doped NiO are obtained within the DFT framework as implemented in QUANTUM ESPRESSO package~\cite{QE}. We used the DFT+U formulation, where the Hubbard $U$ is an on-site correlation energy on the Ni $3d$ orbitals, with the Perdew-Burke-Ernzerhof (PBE) exchange-correlation functional~\cite{PBE} for geometry optimization. A value for the Hubbard $U$ of $U= 4.7$\,eV for PBE+U was used, based on a previous quantum Monte Carlo study for K-doped NiO~\cite{shin2017}. We note that geometrical optimization using PBE+U and variational Monte Carlo simulations give essentially identical results~\cite{wrobel2019}.
For the PBE+U calculations, norm-conserving energy-consistent pseudopotentials proposed by Burkatzki, Filippi, and Dolg were used with 400\,Ry plane-wave kinetic energy cut-off~\cite{burkatzki07,burkatzki08}. In order to describe the 3\% and 6\% doped NiO system in this study, 64 and 32 atoms simulation cells were considered, respectively. These geometries were fully relaxed below to below 10$^{-2}$\,eV/\AA\ Hellman-Feynman forces using 2$\times$2$\times$2 (64 atoms) and 4$\times$4$\times$4 (32 atoms) $\mathbf{k}$-point grids. We also performed one analogous PBU+U calculation on a 32-atom NiO cell with one Li substituted for Ni (6\% doping) using an USPP Li pseudopotential and a 60~Ry kinetic energy cut-off. Atomic positions were relaxed until forces were less than 0.03 eV/{\AA} using $6\times6\times6$ $\mathbf{k}$-point-grid.

To study correlated electronic structures, we adopted the charge-self-consistent DFT+DMFT method as implemented in the Wien2k+eDMFT package~\cite{PRB.81.195107}. The Wien2k calculation was performed using the PBE functional~\cite{PBE} with the $\mathbf{k}-$point grid of 21$\times$21$\times$21 for bulk NiO, 3$\times$5$\times$5 for both 6\% electron- and hole-doped NiO, and 2$\times$2$\times$2 for 3\% hole-doped NiO calculations. Based on DFT band structures obtained from Wien2k, we performed DMFT calculations using the hybridization energy window from -10\,eV to 10\,eV with respect to the DFT Fermi energy for all structures. The same $\mathbf{k}-$point grid as the Wien2k DFT one was used for the DMFT calculation in each structure. The DMFT impurity problem for treating the local correlation effect of Ni $d$ orbitals was solved using the continuous-time quantum Monte Carlo method~\cite{PRB.75.155113,RMP.83.349}. A Coulomb interaction $U=10$\,eV and a Hund's coupling $J=0.9$\,eV were used for interaction parameters. This large value of $U$ is necessary for treating $d-$orbitals in covalent oxides when the wide hybridization window ($\sim$20\,eV) is used for constructing highly localized $d-$orbitals~\cite{PRB.90.075136}. 
For the double-counting potential, we use the so-called nominal double-counting, which is close to the exact double-counting calculation~\cite{PRL.115.196403}.

\section{Results and discussion}

\subsection{Structural analysis of doped and undoped NiO thin films}

\begin{figure}
\includegraphics[width=8.6cm]{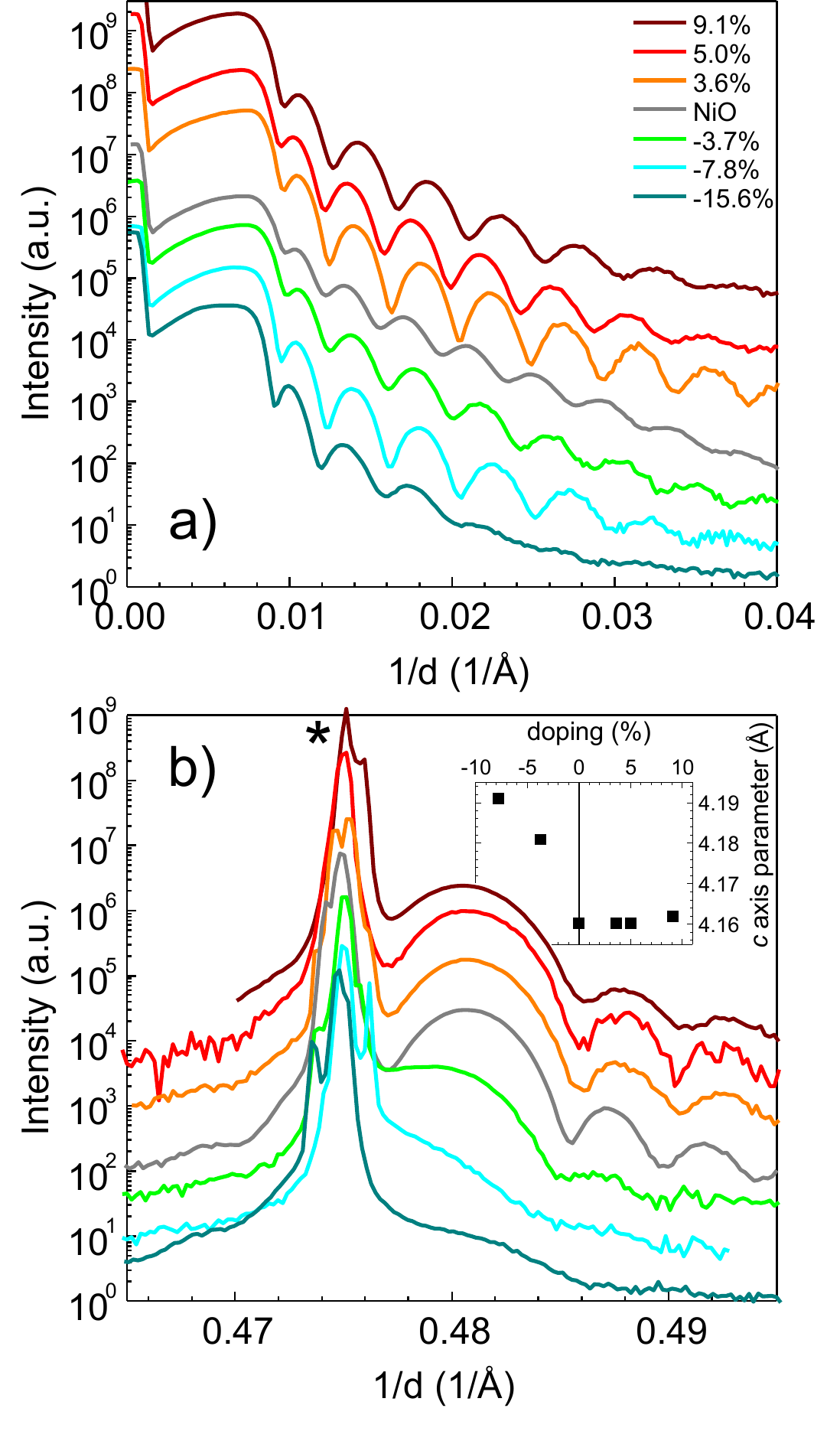}
\centering
\caption{\label{fig:XRD} X-ray diffraction of pure, potassium-doped, and indium-doped NiO thin films with the doping level denoted in the legend. Positive numbers indicate hole- (potassium-)doping and negative numbers electron- (indium-)doping. X-ray intensity is plotted as a function of $1/d = 2 \sin{\theta}/ \lambda $, where $d$ is the spacing between lattice planes, $\theta$ is half the angle between incoming and outgoing beam, and $\lambda$ is the wavelength. (a) Reflectivity, (b) scan through the (002) reflection of substrate and film. The substrate peak is marked with an asterisk and the $c$ axis parameter calculated from the (002) peak position as a function of doping is shown in the inset.}
\end{figure}

Both potassium\cite{wrobel2019} and indium (see Appendix) dope substitutionally on Ni sites. Doped and undoped NiO thin films have smooth interfaces and surfaces as evidenced by the observation of oscillations in x-ray reflectivity measurements, i.e., Kiessig fringes, in Fig.~\ref{fig:XRD}a. The film thickness is the inverse of the spacing between the fringes and is about 50 unit cells for all films. See Appendix for detailed information on film thickness.

Fig.~\ref{fig:XRD}b shows scans through the (002) reflection of substrate (marked with an asterisk) and film. As in the reflectivity from Fig.~\ref{fig:XRD}a, we observe oscillations (in this case Laue fringes) which indicate smooth interfaces and surfaces. The lattice parameters were extracted from the position of the (002) film peak adjacent to the sharp substrate peak. They are displayed in the inset as a function of doping level. We note that the lattice parameter increases much more rapidly with indium- than with potassium-doping. We attribute this different response to indium substitutionally doping NiO and thereby expanding the lattice -- the ionic radius of In$^{3+}$ is about 10\%
~larger than the ionic radius of Ni$^{2+}$ -- whereas in addition to substitution, potassium/oxygen-vacancy clusters are formed upon potassium doping~\cite{wrobel2019}. These defect  clusters can absorb strain locally, which reduces the tendency to a global lattice expansion.  For a detailed analysis and discussion of the structure of the K-doped NiO films, see Ref.~[\onlinecite{wrobel2019}].   

\subsection{Electron energy loss spectroscopy}
\label{subsec:EELS}

\begin{figure}
\includegraphics[width=8.6cm]{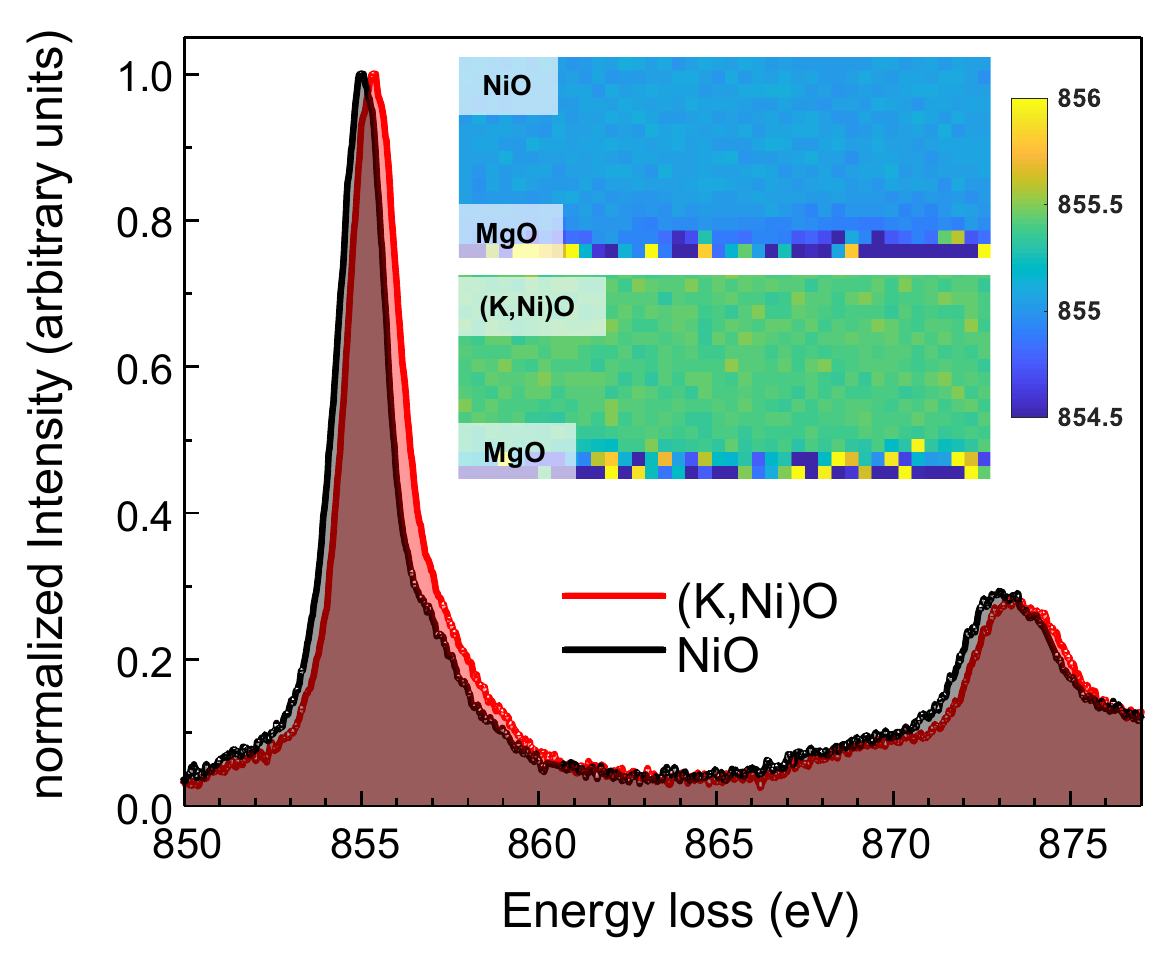}
\centering
\caption{\label{fig:NiEELS} EELS at the Ni $L_{2,3}$-edges of NiO (black line) and 3.6\%-potassium-doped NiO (red line). The inset shows position mapping of the Ni $L_3$-peak over 400\,nm$^2$. This clearly shows that the samples are very homogeneous and that the $L_3$-peak position is shifted towards higher energies for the potassium-doped sample, indicating a higher oxidation state.} 
\end{figure}

\begin{figure}
\includegraphics[width=8.6cm]{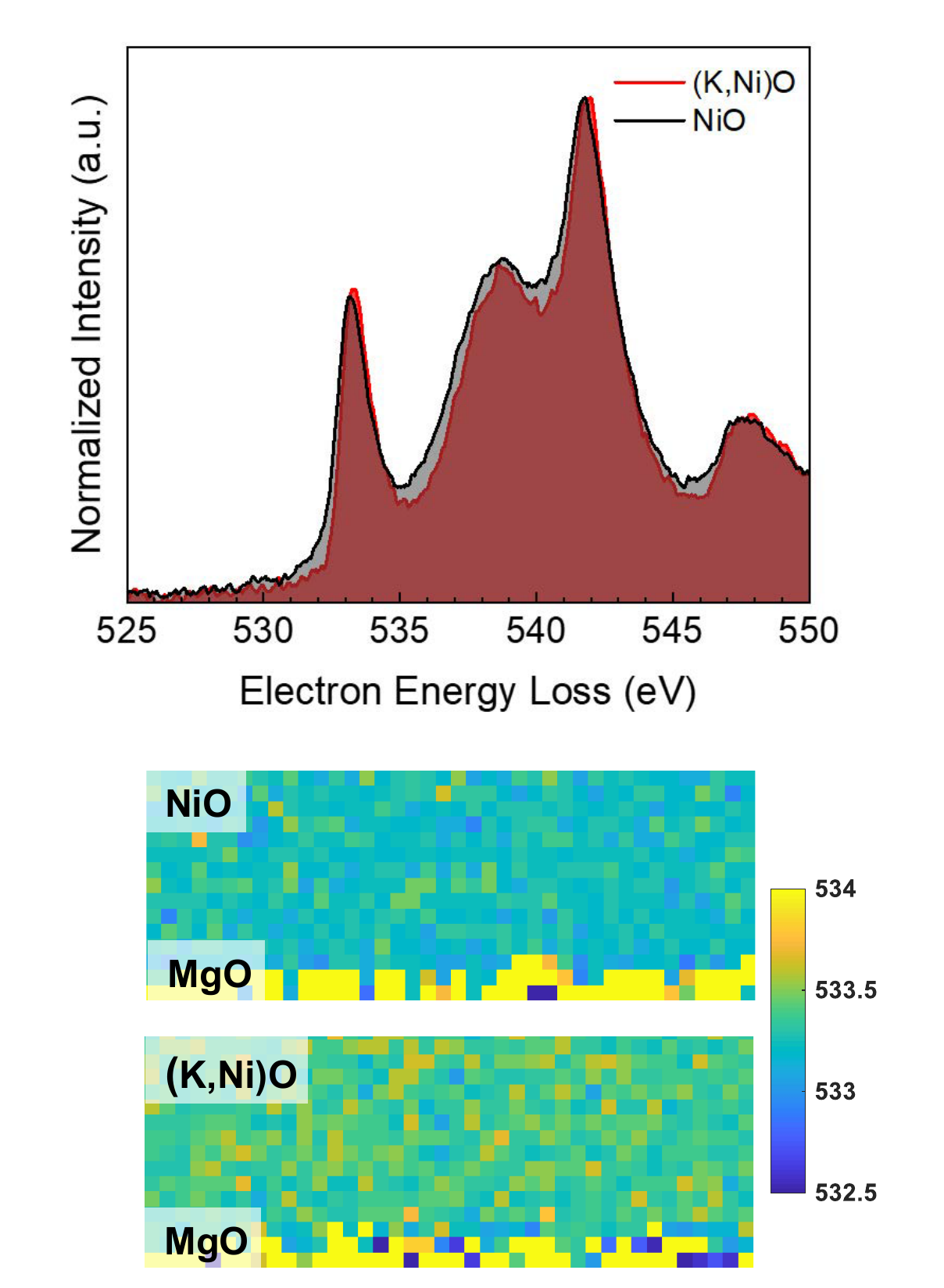}
\centering
\caption{\label{fig:OEELS} EELS at the oxygen $K$-edge of NiO (black line) and 3.6\%-potassium-doped NiO (red line). The lower panel shows position mapping of the O prepeak at about 533\,eV over 400\,nm$^2$. This shows that the prepeak position is slightly shifted towards higher energies for the potassium-doped sample.} 
\end{figure}

Fig.~\ref{fig:NiEELS} and Fig.~\ref{fig:OEELS} show EELS at the nickel $L_{2,3}$-edges and oxygen $K$-edge for undoped and potassium-doped NiO. The spectra are averages over large sample volumes that contain  $\sim 83\times 10^3$ -- $\sim 35 \times 10^4$ unit cells of NiO or (K,Ni)O. In addition, we sampled various smaller spots across both samples but could not find any difference between different areas as shown in the inset of Fig.~\ref{fig:NiEELS} and in the bottom panel of Fig.~\ref{fig:OEELS}.

The mean position of the oxygen prepeak derived from the various spots is 533.25$\pm$0.10\,eV for undoped NiO and 533.40$\pm$0.11\,eV for potassium-doped NiO, which constitutes a small blue-shift which is smaller than the standard deviation. The mean position of the Ni $L_3$-peak is 855.06$\pm$0.03\,eV for undoped NiO and 855.43$\pm$0.03\,eV for potassium-doped NiO. This corresponds to a blue-shift of 0.37\,eV for potassium-doped NiO, which is a significant shift and consistent with an increase in oxidation state~\cite{koyama2005}. Moreover, Ni$^{3+}$ x-ray absorption  spectra show a double-peak structure at the $L_3$-edge~\cite{koyama2005}.  Therefore, for a compound with a slightly larger oxidation state than +II, one would expect a broadening of the $L_3$-peak, which is indeed what we observe. These findings lead us to conclude that the potassium dopants are distributed uniformly throughout the sample (Fig.~\ref{fig:NiEELS}), and that the potassium (hole) doping leads to an increased Ni oxidation state.

We note that, in contrast to EELS\cite{reinert1995} and BIS\cite{kuiper1989character,van1992,hufner1992acceptors} spectra of Li-doped NiO, we do {\em not} see any evidence of a pre-peak at about 529~eV. As we explain later, this pre-peak is absent in K-doped NiO. In Li-doped NiO, this pre-peak appears because defect states get pushed into the gap by a Jahn-Teller distortion of the oxygen octahedra about the Li defect\cite{chen2012}. In K-doped NiO (at low or moderate doping), such a distortion does not take place, and the oxygen octahedra retain their symmetric shape about the K dopant (see Appendix for calculated dopant-oxygen bond lengths in K-doped, In-doped, and Li-doped NiO). We believe that as a consequence, the defect states remain below the Fermi energy.

\subsection{Optical conductivity and electronic structure of undoped NiO}
\label{subsec:Optical}

\begin{figure*}
\includegraphics[width=17.9cm]{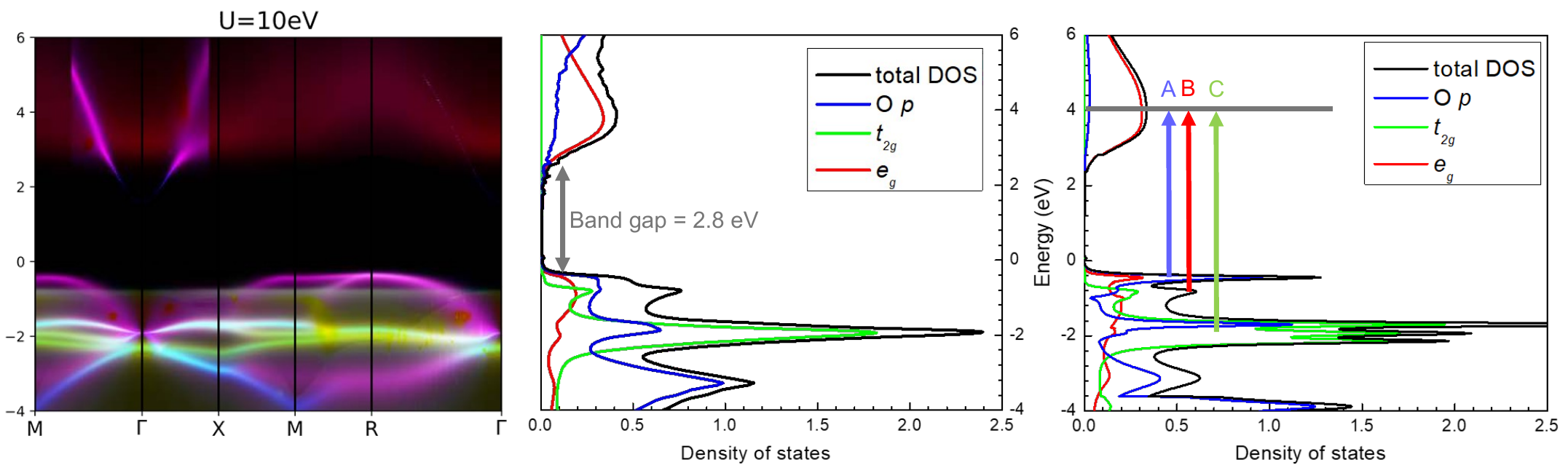}
\centering
\caption{\label{fig:NiODMFT} Left panel: Orbital-resolved DFT+DMFT band structure with $U$ = 10\,eV and $J$ = 0.9\,eV. Different color means different orbital character, i.e., Ni $e_g$ (red), Ni $t_{2g}$ (green), and O $p$ (blue); purple indicates hybridization between O $2p$ (red) and Ni $e_g$ bands, and white indicates hybridization of all three orbital characters.  Middle panel: Total DOS averaged over $k-$points in the whole Brillouin zone. Right  panel: DOS at the M-point where most bands are flat.}
\end{figure*}

\begin{figure}
\includegraphics[width=8.6cm]{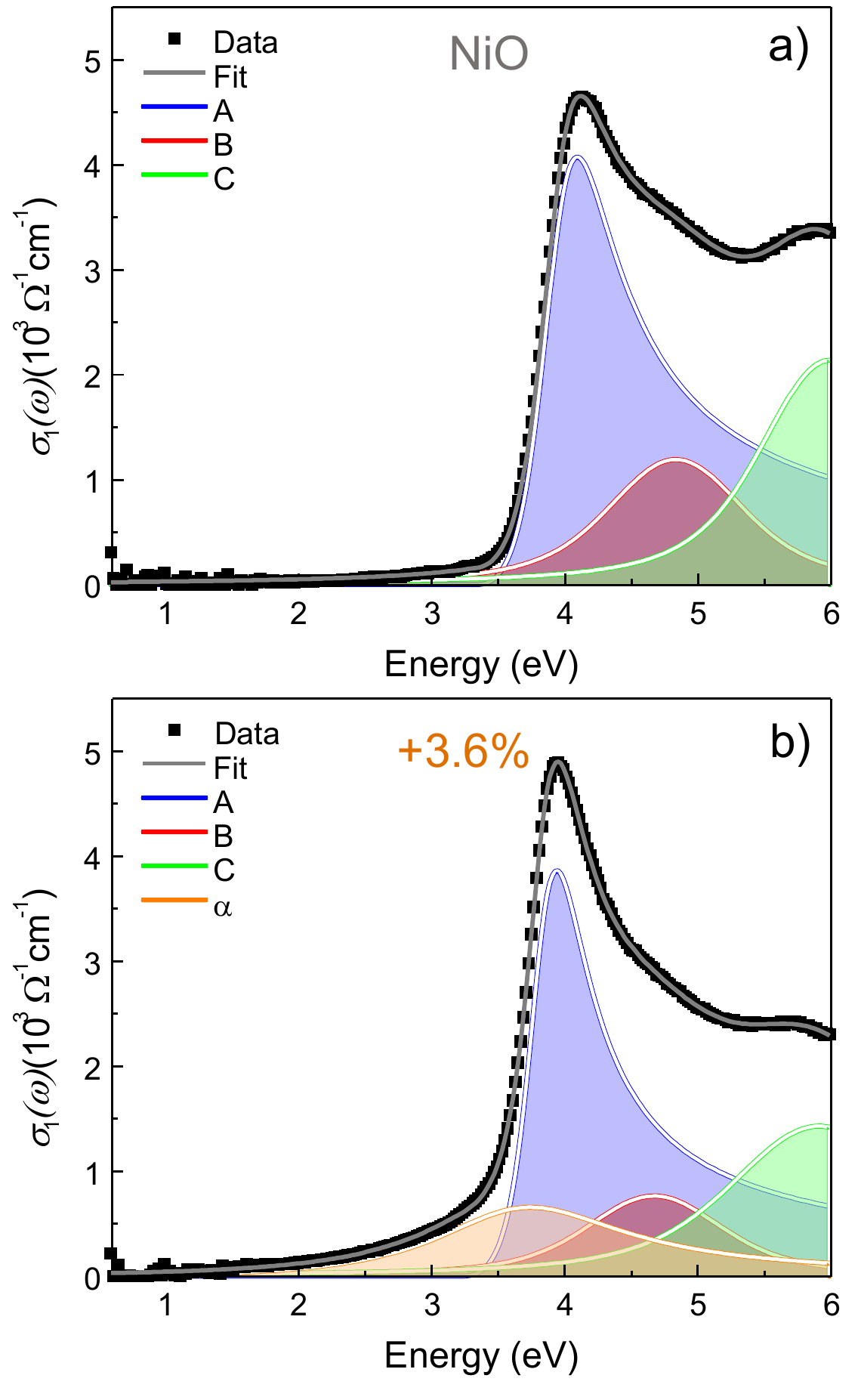}
\centering
\caption{\label{fig:Fits} Optical conductivity of NiO (a) and 3.6\% potassium doped (b) films and fits to the data as indicated in the legend.}
\end{figure}

We start by comparing the DMFT band structure and density-of-states (DOS) (Fig.~\ref{fig:NiODMFT}) of undoped NiO with the measured optical conductivity $\sigma(\omega)$ (Fig.~\ref{fig:Fits}a). In the measured energy range between 0.6\,eV and 6\,eV, there are three distinct peaks, which we name in alphabetic order A, B, and C (from low to high energies). For a quantitative comparison, we fitted $\sigma(\omega)$ with one Tauc-Lorentz and two Voigt models. Peaks in spectroscopy measurements are typically fit with either a Lorentz oscillator (Eq.~\ref{eq:LO})~\cite{wooton1972}, a Tauc-Lorentz model (Eq.~\ref{eq:TL})~\cite{jellison1996}, or a Voigt profile (Eq.~\ref{eq:V})~\cite{herman1996}.

The Lorentz oscillator model is derived from a classical model of non-interacting atoms that describes single, direct interband transitions occurring in insulators~\cite{wooton1972}:

\begin{equation}\label{eq:LO}
    \sigma_{LO}(\omega) = \frac{\gamma_{LO} \, A_{LO} \, \omega^2}{(\omega^2 - \omega_{LO}^2)^2 + \gamma_{LO}^2 \, \omega^2}
\end{equation}

Here, $\gamma_{LO}$ corresponds to the scattering rate, which originates from processes that are referred to as homogeneous broadening, for example, spontaneous emission and collisions~\cite{herman1996}. $A_{LO}$ corresponds to the strength, and $\omega_{LO}$ to the resonance frequency of the Lorentz oscillator. This model works particularly well for, e.g., isolated features in solids, like dopants and excitonic resonances~\cite{herman1996}.

The Lorentz oscillator model does not describe well cases where interacting atoms and electrons dominate, i.e., most solids. In particular, it does not capture the onset of absorption above the band gap. A better approach in these cases is the Tauc-Lorentz profile (Eq.~\ref{eq:TL}). It is derived from a multiplication of the Tauc joint density of states (Eq.~\ref{eq:T})~\cite{tauc1966}, which introduces the onset of absorption, with the Lorentz oscillator model~\cite{jellison1996}:

\begin{equation}\label{eq:T}
    \sigma_{T}(\omega) = \begin{cases} \dfrac{A_{T} \, (\omega - \omega_{gap})^2 }{\omega} &, \omega > \omega_{gap} \\
    0 &, \omega \leq \omega_{gap} \end{cases}
\end{equation}

Here, $A_T$ is the amplitude factor for the Tauc joint density of states and $\omega_{gap}$ is the optical band gap. The Tauc-Lorentz model is then described by the following equation:

\begin{equation}\label{eq:TL}
    \sigma_{TL}(\omega) = \begin{cases} \dfrac{A_{TL} \, (\omega - \omega_{gap})^2 }{(\omega^2 - \omega_{TL}^2)^2 + \gamma_{TL}^2 \omega^2} &, \omega > \omega_{gap} \\
    0 &, \omega \leq \omega_{gap} \end{cases}
\end{equation}

where $A_{TL}$ is the amplitude, $\omega_{TL}$ is the peak position, and $\gamma_{TL}$ is the peak width. This model captures the sharp onset of absorption above band gap and has been used successfully in particular for amorphous or nano/microcrystalline semiconductors~\cite{tompkins2005}.

Another reason for a deviation of the Lorentzian line shape is inhomogeneous broadening caused by, e.g., thermal motion and defects. Inhomogeneous broadening leads to a Gaussian peak shape. When homogeneous and inhomogeneous broadening are of a similar magnitude, the convolution of Gaussian and Lorentzian, i.e., the Voigt profile (Eq.~\ref{eq:V}) may be used~\cite{herman1996}:

\begin{gather}\label{eq:V}
    \sigma_{V} (\omega; A_V, \omega_V, \sigma_G, \gamma_L) = \frac{A_V \mathrm{Re}[w(z)]}{\sigma_G \sqrt{2\pi}} \\
    \mathrm{where} \quad
    z = \frac{\omega - \omega_V + i\gamma_L}{\sigma_G \sqrt{2}} \\
    \mathrm{and} \quad
    w(z) = \exp{(-z^2)} \, \mathrm{erfc}(-iz)
\end{gather}

Here, $\mathrm{erfc}()$ is the complementary error function, $A_V$ is the peak amplitude and also the spectral weight, $\omega_V$ is the peak position, $\sigma_G$ is the Gaussian width, and $\gamma_L$ is the Lorentzian width.
The Voigt model is a very generalized model that is suitable for most of peaks in spectroscopy because in real, experimental systems, both, homogeneous and inhomogeneous broadening, are usually observed.

We use a Tauc-Lorentz model for peak A to account for the sharp edge, we use two Voigt profiles for peaks B and C, and we constrain $\sigma_G$ to be equal to $\gamma_L$. The overall fit to the spectrum is excellent (Fig.~\ref{fig:Fits}a).

We compare the extracted peak positions with the DMFT calculations shown in Fig.~\ref{fig:NiODMFT}. The left panel shows the DMFT band structure as orbital-resolved spectral functions. Red color corresponds to $e_g$ bands, green to $t_{2g}$, and blue to the remaining bands, which are mostly of oxygen $2p$ character. Purple (light blue) colors correspond to mixtures between $e_g$ ($t_{2g}$) and O $2p$ bands. Optical conductivity measures mainly direct transitions, i.e., without momentum transfer, corresponding to vertical transitions in the band structure in Fig.~\ref{fig:NiODMFT}. The intensity of optical conductivity is a convolution of matrix elements with densities-of-states (DOS). Parallel regions (in energy) of the band structure have large DOS and constant energy separation, and therefore contribute strongest to the spectrum. In our case, all bands in the energy range of interest are mostly flat (and therefore parallel) around the M- and R-points, whereas they are highly dispersive and not parallel around the $\Gamma$-point. In particular, we find only one weakly dispersing, empty band (in red, corresponding to mostly $e_g$ character, around 4\,eV), and we assume that the spectral weight in the optical conductivity mostly stems from transitions into this band. To gain insight into the nature of transitions, we look at the DOS at the M-point. Any peak in the occupied DOS at the M-point should correspond to a peak in optical conductivity. Indeed, there is peak just below the Fermi energy which has mostly oxygen $2p$ character, another around -1\,eV with mixed character, and a third one at about -2\,eV with mostly $t_{2g}$ character, which splits up into three different bands depending on momentum. Due to the dispersion of the bands and the resolution of the measurement, the band splitting is not observed in optical conductivity, but the corresponding transition appears as one peak at around 6\,eV. Overall, we observe three peaks in the optical conductivity at about 4~ eV, 5~eV, and 6~eV, consistent with peaks in the occupied DOS. The energy differences between the unoccupied $e_g$ states and the three peaks in the occupied DOS agree quantitatively and also rather well quantitatively with the measured peak energies in the optical conductivity. We can therefore identify peak A as a transition from the mainly O $2p$ valence band, peak B as a transition from the mainly Ni $e_g$ band, and peak C as a transition from the Ni $t_{2g}$ band to the empty $e_g$ band. However, we note that all states are strongly hybridized, in particular the states giving rise to peak B, and the band characterization we use here is just a label for convenience.

\subsection{Evolution of the electronic structure with doping}
\label{subsec:Evolution}

\begin{figure}
    \includegraphics[width=8.6cm]{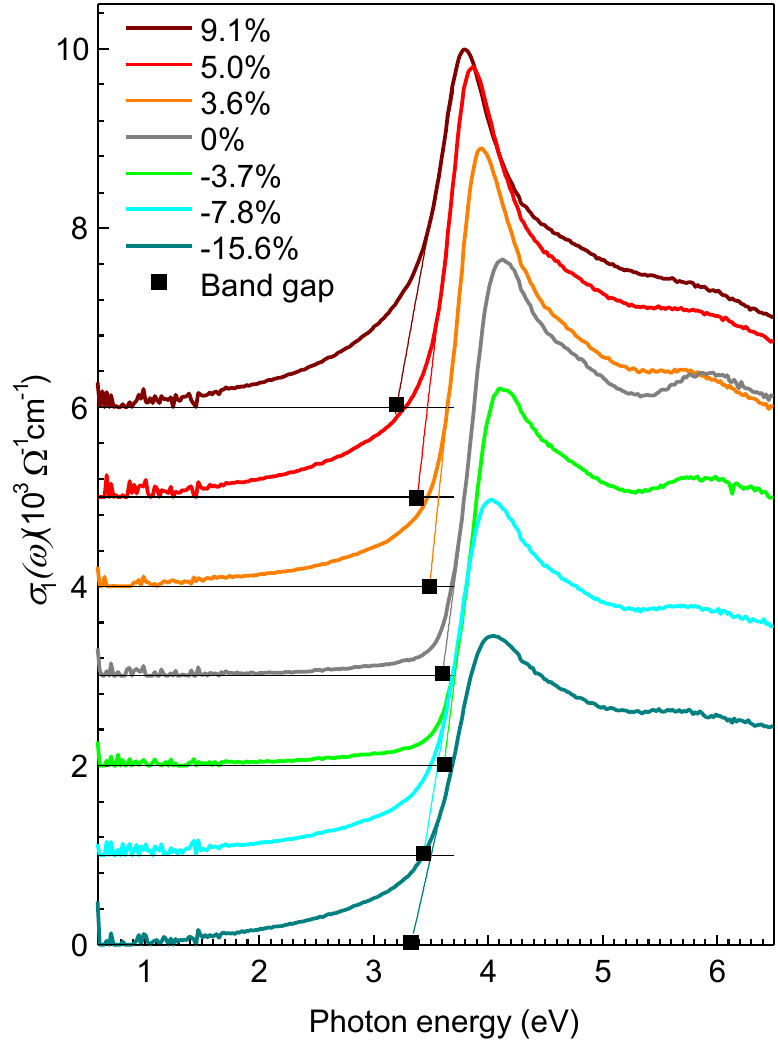}
    \centering
    \caption{\label{fig:S1} Evolution of the optical conductivity $\sigma_1(\omega)$ with hole and electron doping, as denoted in the legend. Positive (negative) numbers correspond to potassium (indium) doping. The curves are offset for clarity. The black squares indicate the optical band gap obtained from extrapolation of the absorption edge to zero (indicated by solid lines) for different doping levels.}
\end{figure}

\begin{figure}
    \centering
    \includegraphics[width=8.6cm]{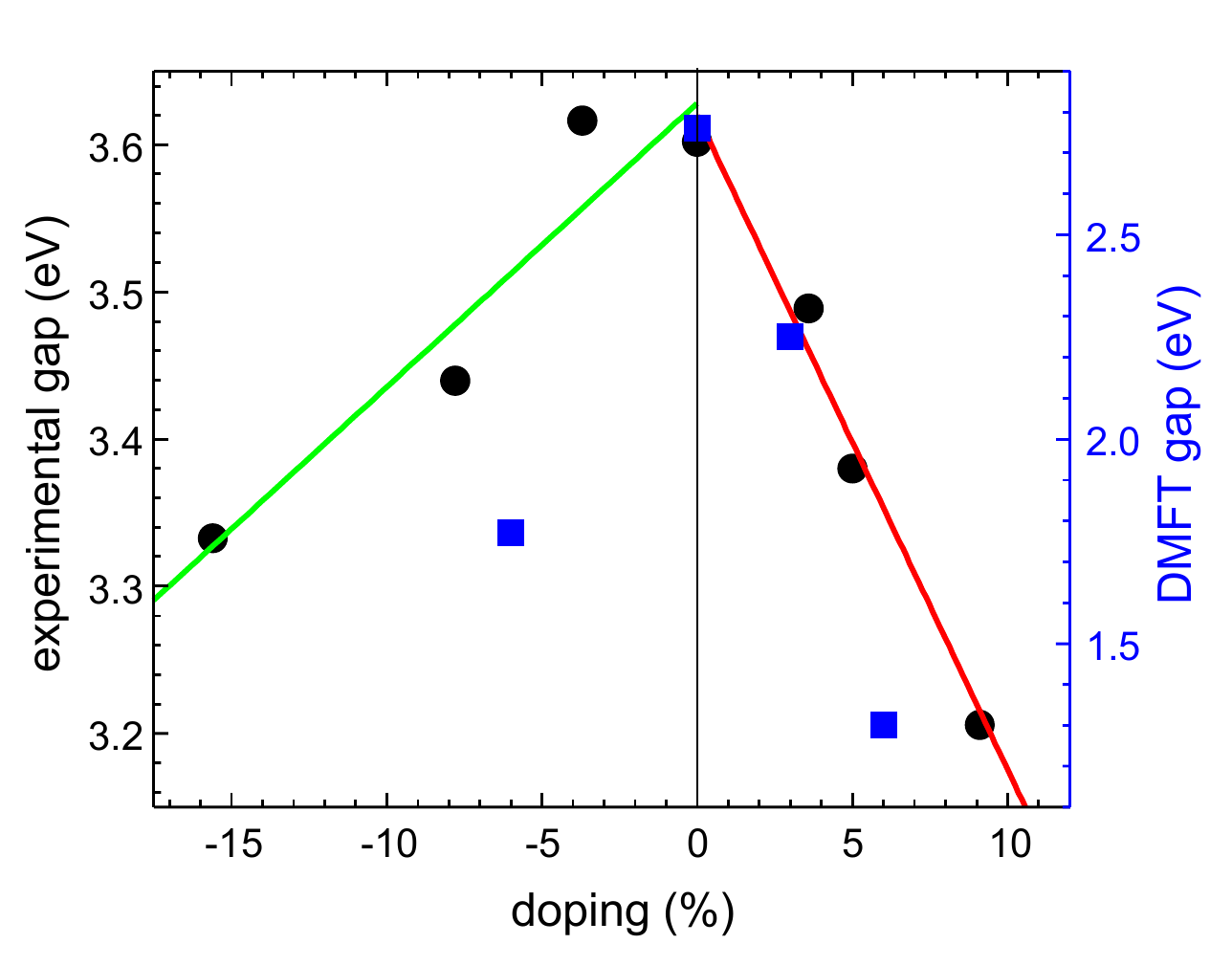}
    \caption{Band gap as function of dopant concentration, with negative (positive) for electron (hole) doping extracted from the optical conductivity spectra (circle dots; Fig.~\ref{fig:S1}) and the DMFT DOS (square dots; Fig.~\ref{fig:DOS_doping}).}
    \label{fig:band_gap}
\end{figure}

\begin{figure}
\includegraphics[width=8.6cm]{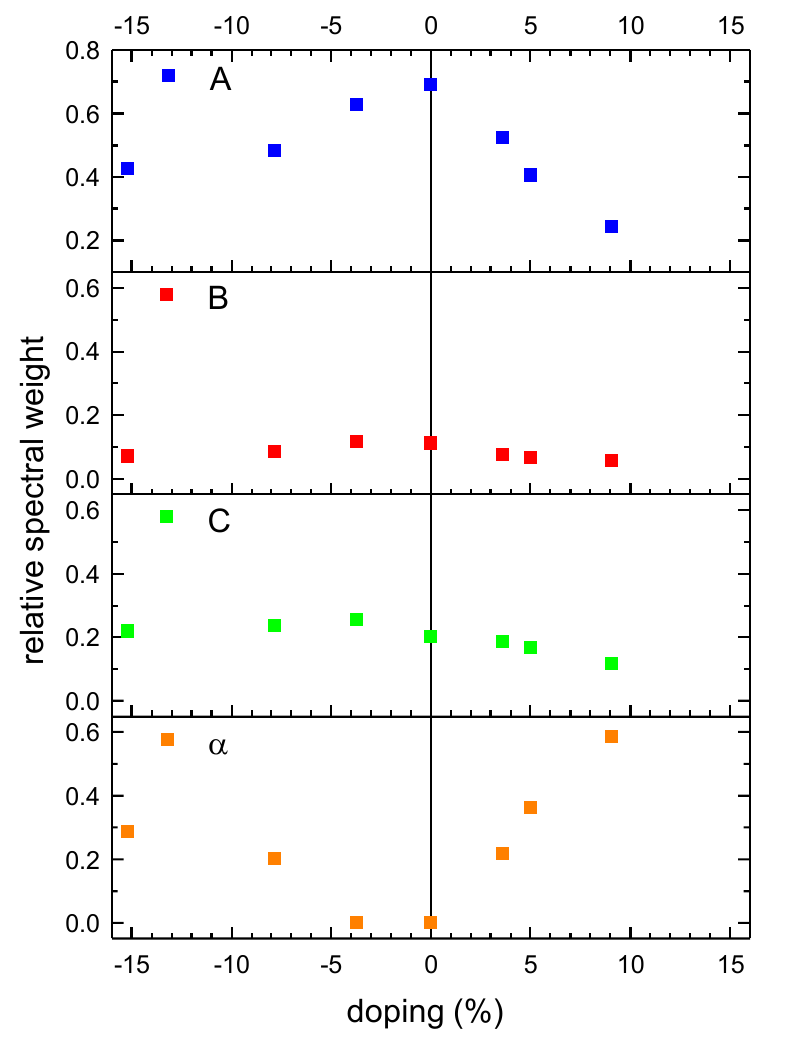}
\centering
\caption{\label{fig:weight_doping} Relative spectral weight with doping, which corresponds to the peak area divided by the sum of all four peak areas. The widths of the Voigt peaks was restricted so that they would not significantly contribute to the spectral weight below the gap. The loss in spectral weight of peak A is almost exclusively compensated by a gain in spectral weight of the Lorentz oscillator peak $\alpha$, i.e., we observe a transfer of spectral weight from A to $\alpha$.}
\end{figure}

After discussing details of the band structure of undoped NiO and the relation to its optical spectrum, we now turn towards hole (= potassium) and electron (= indium) doped NiO. The measured optical conductivity at various levels of doping is shown in Fig.~\ref{fig:S1}. We note that for both electron and hole doping, the optical spectra evolve continuously with doping. In general, the main peak shifts towards lower energy, the spectral weights of peaks decrease, and the low-energy tail below the main peak increases in weight. Fig.~\ref{fig:weight_doping} shows the evolution of the relative spectral weight, i.e., peak area divided by sum of the area of all fitted peaks, with doping.

\begin{figure}
    \includegraphics[width=8.6cm]{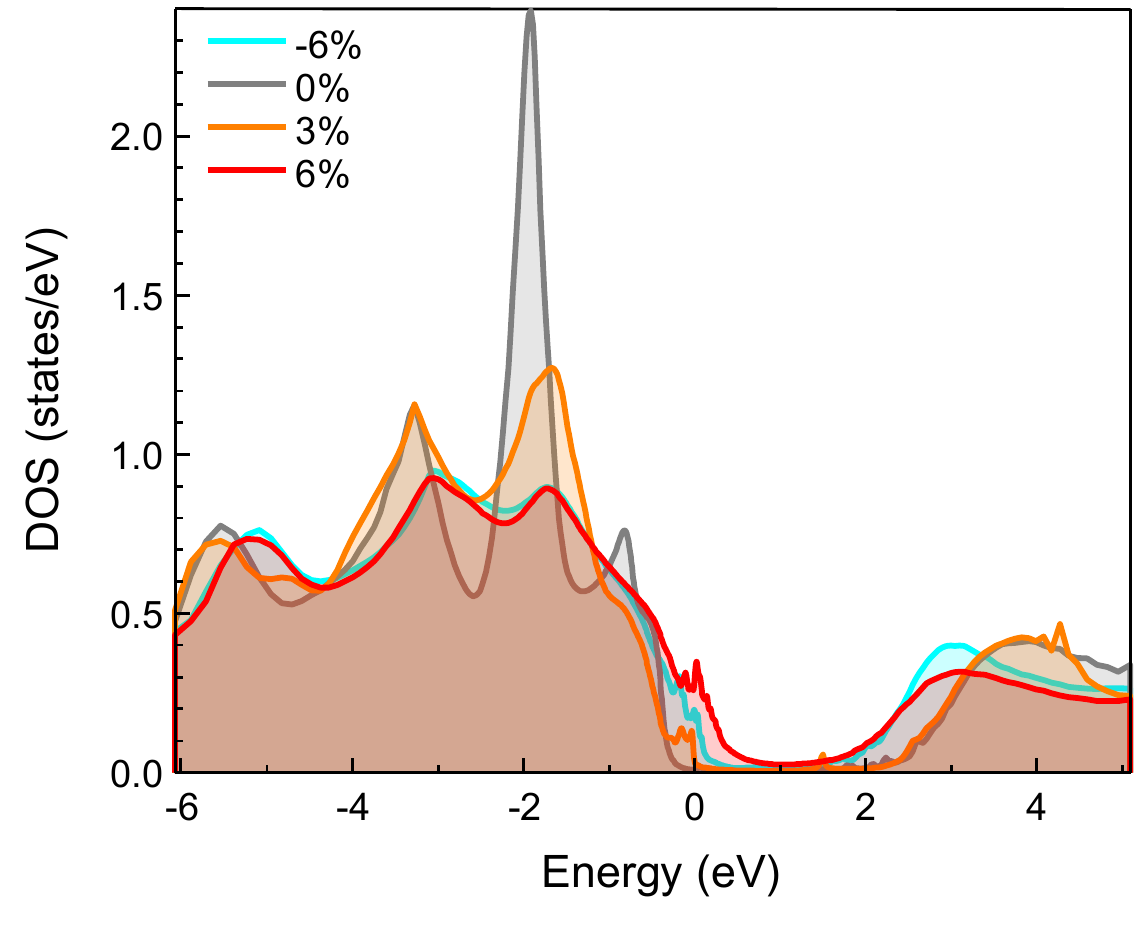}
    \centering
    \caption{\label{fig:DOS_doping}Evolution of DOS with doping. Blue color and negative number indicate electron (=indium) doping and red/orange colors and positive numbers indicate hole (=potassium) doping.}
\end{figure}

We find that fitting the spectra of doped NiO requires the addition of a fourth peak, peak $\alpha$, with a Lorentzian line shape, as shown in Fig.~\ref{fig:Fits}(b). As mentioned above, the Lorentzian oscillator model is derived from a classical model of non-interacting particles and therefore is suitable for isolated dopants in solids.
As was the case for undoped NiO, the fits to the spectra are excellent. In the following paragraphs, we discuss the evolution of different features and how they relate to the evolution of the DOS (Fig.~\ref{fig:DOS_doping}).

\begin{figure}
    \includegraphics[width=8.6cm]{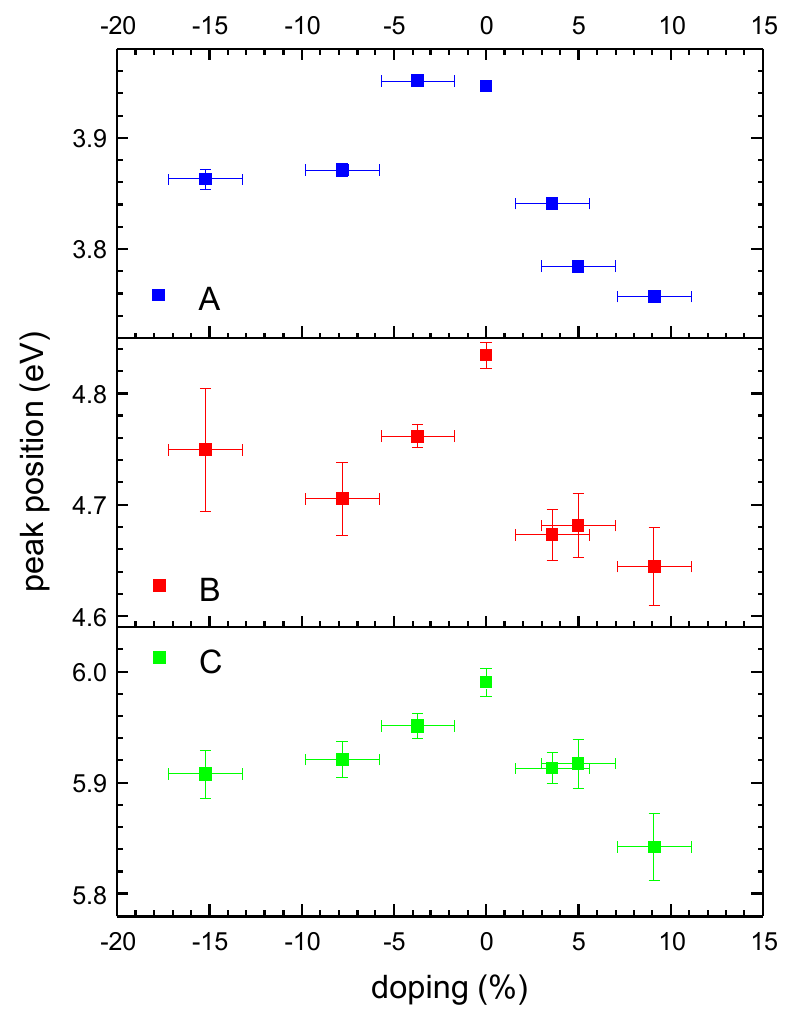}
    \centering
    \caption{\label{fig:peak} Peak positions of the Tauc-Lorentz, Voigt1, and Voigt2 peak as a function of doping level. The error in doping level is estimated to be $\pm$2\%\ based on typical errors in RBS analysis and the errors in peak position result from the fitting procedure.}
\end{figure}

With both, hole- and electron-doping, the positions of the three peaks (A, B, and C) move towards lower energies. This behavior is visible to the bare eye in Fig.~\ref{fig:S1} for peak A at about 4\,eV. The two other peaks (B and C) move at the same rate (energy shift vs. doping), as can be seen in Fig.~\ref{fig:peak}, which is the result of a peak fitting procedure as shown in Fig.~\ref{fig:Fits}(b) for potassium doping of 3.6\%. Overall, the shift is larger with hole- than with electron-doping. 

The fourth peak $\alpha$ is very broad, which leads to ambiguity and large error bars in determining the peak position; we will therefore refrain from attempting to put the position of the $\alpha$ peak in the context of the experimental data. In practice, it is mainly determined by the constraint implemented in the fitting procedure that the peak position of peak $\alpha$ has to be below that of peak A. 

The experimentally measured peak shifts cannot be directly and quantitatively compared to the calculated DMFT DOS because the momentum-resolved DOS is unattainable for the following reason. The DMFT calculations for doped NiO have to be run in a larger unit cell, which, for 6\%\ doping, is 16 times larger than the NiO unit cell and, for 3\%\ doping, 32 times times larger. This corresponds to 1 out of 16 (32) nickel sites being replaced by In or K to achieve a 6\%\ (3\%) doping level. It is not feasible to calculate the band structure for such large unit cell sizes. Therefore, we cannot track the changes in band structure with doping at, e.g., the M-point. For the same reason, a distinction between $e_g$ and $t_{2g}$ levels in the partial DOS becomes ambiguous. Therefore, we instead qualitatively compare the evolution of the total DOS with the evolution of the optical spectrum.

The spectral weight in optical conductivity is a result of the number of states electrons occupy and the number of empty states electrons can be excited into (as well as the amplitude of matrix elements). Hence, when looking at the evolution of amplitude with respect to doping and comparing with the DOS, we need to take both aspects into account. For both hole and electron doping, we observe a decrease in amplitude of all three initial peaks (A-C). The overall decrease may be explained by a decrease of available empty states $e_g$, which is revealed in the DOS shown in Fig.~\ref{fig:DOS_doping}. This decrease is stronger for hole than for electron doping, which is also in agreement with the DMFT-derived DOS. For hole doping, a de-population of the valence band O $2p$ and Ni $e_g$ states, which have rather narrow DOS will also lead to a rapid decrease in spectral weight.

For further analysis, it is worthwhile to look at the relative spectral weight, i.e., the peak amplitude of each peak divided by the sum of the amplitudes of all three or four peaks, as shown in Fig.~\ref{fig:weight_doping}. The relative spectral weight of peaks B and C remain relatively constant, indicating that the absolute decrease of peaks B and C is due to the decrease in the unoccupied DOS.

As noted earlier, we observe the appearance of a fourth contribution, peak $\alpha$, upon both electron and hole doping. This peak is centered in the gap region just below peak A, and its amplitude increases at approximately the same rate as the amplitude of peak A decreases. The amplitude of peak $\alpha$ grows faster with hole doping than with electron doping. This finding suggests that a spectral weight transfer occurs from the undisturbed O 2$p$ band to new, localized states that form in the vicinity of the dopant due to changes in the chemical bond. This is consistent with the general picture of the evolution of spectra in correlated systems with doping~\cite{reinert1995,kunevs2007prl,shinohara2015}, in which spectral weight is transferred from the conduction band towards the gap. 
This is also consistent with DFT+U  calculations showing that the charge introduced by an indium dopant is distributed over the neighboring nickel and oxygen sites, forming localized states rather than modifying the existing oxygen $p$-band (see Appendix for information).

Figure~\ref{fig:DOS_doping} shows the calculated DMFT DOS for undoped NiO, 3\% and 6\% K doping, and 6\% In doping. For both electron and hole doping, we observe mostly the same trends with a few exceptions:
for both, electron and hole doping, there appear a narrow band of occupied states at the top of the valence band, which is more pronounced for hole than for electron doping. Moreover, the conduction band Ni $e_g$ peak is reduced in energy and amplitude but stronger so for hole doping. The appearance of the occupied states at the top of the valence band is consistent with and quantitatively in agreement with the appearance of the Lorentz peak in the optical conductivity. The larger amplitude for hole doping of the new DOS peak at the top of the valence band is consistent with the experimentally observed larger relative spectral weight for the Lorentz peak for hole doping than for electron doping (Fig.~\ref{fig:weight_doping}). 

To gain further insight into the electron-doped system, we calculated orbital projections of L\"owdin charges and polarization for $\sim$6\% K- or 
In-doping (one dopant in a 32-atom supercell)  
within the GGA+U scheme (calculated orbital-resolved L\"owdin charges are detailed in the Appendix). While the total L\"owdin charges assigned to an atomic site may not be useful, the change in projections on Ni $d$ orbitals gives some insight. Analysis of the results shows that for In (electron) doping, there is an increase in the $d$-orbital occupancy of the minority-spin $e_g$ orbitals, thus reducing the net spin polarization of these Ni atoms.  The nearest O atoms also attain some spin polarization through a slight increase (decrease) in the population of $p_x$ and $p_y$ orbitals anti-parallel (parallel) to the spin polarization of the removed Ni atom. 
For K (hole) doping, there is a reduction in the occupancy of $d$ states on the nearest-neighbor Ni shell in both spin channels.  

\begin{figure}
    \includegraphics[width=8.6cm]{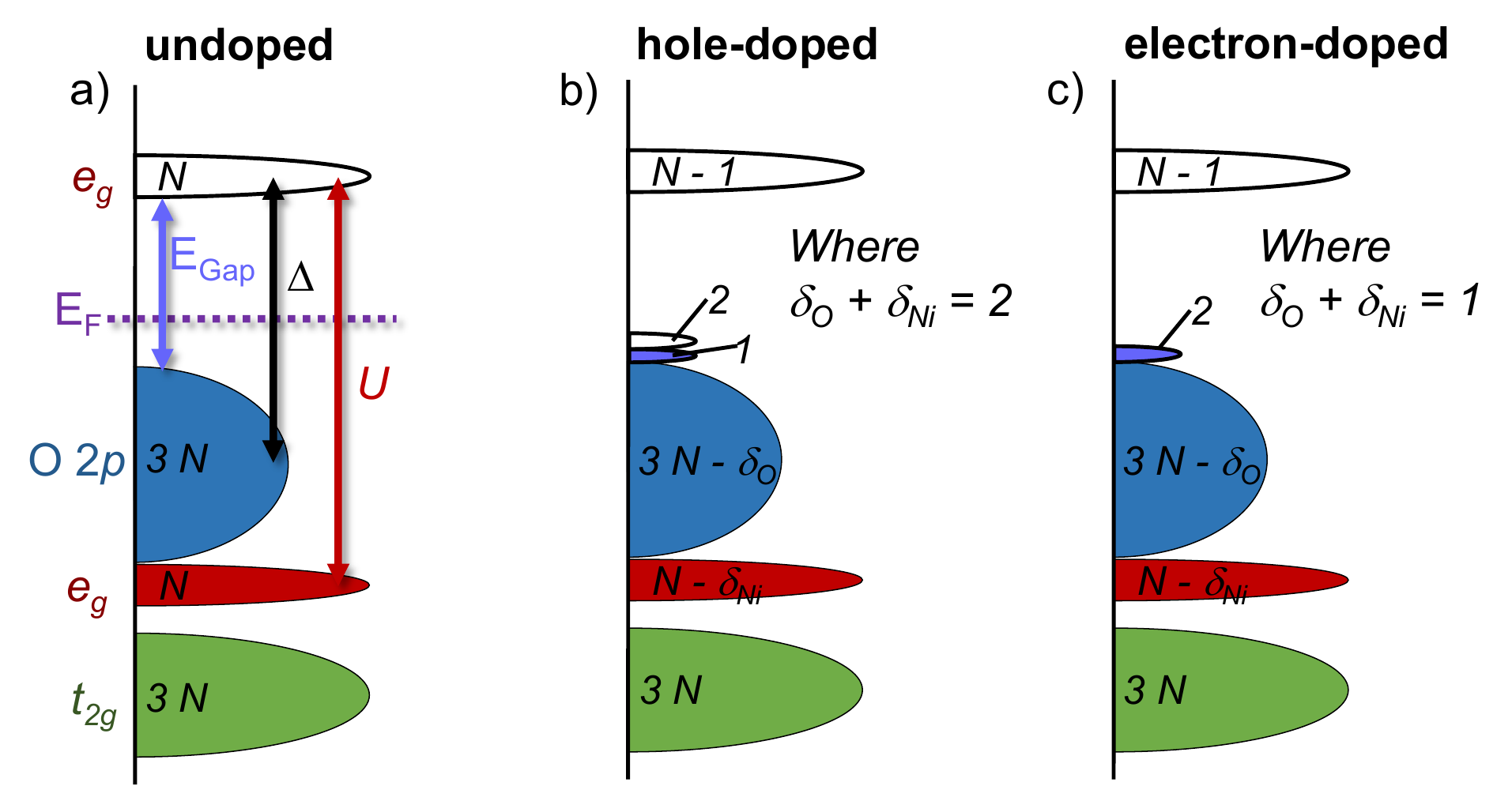}
    \centering
    \caption{\label{fig:DOS_sketch} Cartoon of electron densities of states (DOS) of undoped (a), hole-doped (b), and electron-doped (c) NiO.  
    E$_{\mathrm{F}}$ denotes the Fermi level, E$_{\mathrm{Gap}}$ the band gap, $\Delta$ the charge transfer energy, i.e., the energy needed to transfer an electron from the O 2$p$ band to the upper Hubbard, and $U$ is the Hubbard repulsion.
    In NiO, all six $t_{2g}$ levels, and six O 2$p$ levels are occupied, whereas two out of four $e_g$ levels are occupied, and the separation between occupied and unoccupied $e_g$ levels corresponds to the on-site correlation (Hubbard repulsion) $U$.
    b) Due to the strong hybridization between O 2$p$ and Ni 3$d$ levels, the removal of one electron, i.e., hole-doping, also affects the Ni $e_g$ levels (as opposed to solely creating a hole in the O 2$p$ band as is expected for an archetypical CT insulator). This leads to the creation of three new states close to the Fermi level, one occupied and two unoccupied, and removes spectral weight from the lower and upper Hubbard band as well as from the O 2$p$ band.
    c) Due to the strong hybridization between O 2$p$ and Ni 3$d$ levels, the addition of one electron, i.e., electron-doping, also affects the O 2$p$ levels (as opposed to creating two  occupied states just below the upper Hubbard band at the expense of upper and lower Hubbard band as is expected for an archetypical CT insulator). This leads to the creation of two  states at the Fermi level and removes spectral weight from the lower and upper Hubbard band as well as from the O 2$p$ band.}
\end{figure}

What is surprising is that for {\em both} electron and hole doping, the DMFT DOS shows that there are occupied states appearing near or at the valence band edge, which is consistent with our optical conductivity spectra. Naively, one might expect that hole doping would lead to some unoccupied states near the valence band edge, and electron doping to some occupied states mid-gap or near the conduction band edge.   
Reinert {\em et al.}~\cite{reinert1995} find electron and hole doping give rise to mid-gap states, but in their case, hole-doping gives rise to an empty state 1.2\,eV above the valence band edge, while electron doping (through oxygen deficiency) gives rise to Ni$^0$ defect states 0.6\,eV and 1.7\,eV above the valence band edge, the first of these an occupied  pinning center. Similarly, Oka {\em et al.}~\cite{oka2012} also found an occupied state about 0.8\,eV above the valence band edge. 
The appearance of defect states in Li-doped NiO in the gap, the first of these about 1~eV above the valence band edge, has been attributed to a Jahn-Teller like structural distortion\cite{chen2012}. In GGA+U and HSE06 calculations\cite{chen2012}, it was shown that without relaxation, the defect states remained at the valence band edge, below the Fermi energy. When the system was relaxed, they were pushed above the Fermi level into the gap. Our DFT+U calculations\cite{shin2017} show that oxygen octahedron about the K dopant is very symmetric with no indication of a Jahn-Teller like distortion, and the defect states remain at the valence band edge, just below the Fermi energy. The same result is obtained from DMFT calculations, using relaxed structures from our PBE+U calculations (see Fig.~\ref{fig:DOS_doping}). As a check, we also performed PBE+U calculations on 6\% Li-doped NiO (see Appendix). For the relaxed structure, we obtained defect states in the gap, consistent with the results of Chen and Harding\cite{chen2012}. 
Furthermore, the appearance of mid-gap states for K-doped NiO is 
not consistent with our experimental optical conductivity or DMFT DOS as we see a smooth evolution of the band gap and spectral peaks with dopant concentration, and not distinct energy levels independent of dopant concentration. Finally, we would not expect to have Ni$^0$ states in our system that is electron-doped with In$^{3+}$ rather than through deficiency in O$^{2-}$. Instead, electron doping leads to a state right at the valence-band edge that acts as a pinning center, as is often the case for metallic cations. Fig.~\ref{fig:DOS_sketch} depicts cartoons of the DOS for undoped, hole-doped, and electron-doped NiO.

The behavior we observe under hole doping is qualitatively consistent with that reported by Kunes {\em et al.}~\cite{kunevs2007prb} in that they find a large reduction of spectral weight at a few eV above the gap, and the appearance of a mid-gap peak. However, their results are for relatively large hole concentrations (38\%). Therefore, it is difficult to make any quantitative comparison with our results for low hole doping.

Fig.~\ref{fig:band_gap} shows the experimental band gaps as extracted from Fig.~\ref{fig:S1}, together with estimates of the DMFT band gaps. It is important to note that the DMFT band gaps are estimated by extrapolating the valence band DOS and conduction band DOS from inflection points to zero energy, similarly to what is done for the optical conductivity measurements. This is not an accurate determination of the optical band gaps, but gives a trend in qualitative agreement with the optical conductivity measurements. Both experimental gaps and the estimated DMFT gaps show a marked asymmetry in that the gap reduction is much faster for hole-doping than for electron-doping. This is consistent with the behavior of a charge-transfer insulator; for a Mott insulator, the gap evolution would be symmetric in hole and electron doping.

\section{Conclusion}

We have used a wide range of experimental and theoretical methods to investigate the evolution of the electronic structure of NiO, an archetypical charge-transfer insulator, upon electron and hole doping. We find that the band gap evolves very smoothly under both electron and hole doping, with no indication of defect states, either occupied or unoccupied, that are well separated from the valence band or conduction band edges (to within our experimental resolution). Analysis of optical conductivity spectra show that doping leads to the emergence of a new peak, just below the main peak that stems from the main direct optical transition. 
The emergence of this new peak and its position are in agreement with electronic DOS calculated\cite{shin2017} using DFT+U and the dynamical mean field method, which show that for both electron and hole doping, occupied states appear right at the valence band edge. The appearance of defect states at the valence band edge is different from the behavior observed in Li-doped NiO, in which a Jahn-Teller like distortion of the oxygen octahedron\cite{chen2012} about the Li dopant pushes the defect states above the valence band edge into the gap above the Fermi level\cite{kuiper1989character,van1992,hufner1992acceptors,reinert1995}. The effect of the distortion in Li-doped NiO on the electronic states is in conceptual agreement with studies showing that structural distortions in correlated materials have significant effects on the electronic structure and defect levels\cite{biermann2005,giovannetti2014,vasudevan2018}.  
The observed shift in the peak position of the main peak (peak A) in the optical conductivity is also consistent with changes in the DMFT DOS. 
The evolution of the experimental optical band gap is asymmetric with respect to electron and hole doping, with the gap diminishing more rapidly with hole than with electron doping. This asymmetry is consistent with the behavior of the calculated DOS and the gap estimated from the DOS, and is indicative of a charge transfer insulator rather than a Mott insulator. 
%
%
We observe a reduction in spectral weight of the optical transitions from the valence band to the conduction band, with an increase in spectral weight of a transition from the defect states at the top of the valence band to the conduction band. Again, this evolution is very smooth with dopant concentration, and we did not reach concentration levels at which the spectral weights of the valence to conduction band transitions collapse.

Analysis of the experimental and theoretical results presents the following picture.  Hole doping with K gives rise to a shallow acceptor level very close to the valence band edge (see Fig.~\ref{fig:DOS_sketch}b), consistent with high p-type conduction typically observed in NiO due to Ni deficiency. Because the valence edge is predominantly of O $2p$ type, with some spectral-weight from Ni $e_g$  states hybridized with O $2p$, electrons are depleted from the $p$-$d$ hybridized states close to the valence band-edge.  Holes can then be donated to the valence band, giving rise to p-type conduction. 
We note that the hole density resides mostly on the K-dopant\cite{shin2017}, while in Li-doped NiO it resides on next-nearest neighbor Ni\cite{chen2012}.  
On the other hand, electron doping with In gives rise to deep donor traps, consistent with the absence of any n-type conduction in pure NiO.  The deep donor trap states form because the excess electrons from In goes to predominantly a spatially localized (near In) Ni $e_g$ up- and down-spin states, i.e. changing both the up- and down-spin densities, effectively reducing their on-site repulsion (the Hubbard $U$). Spatial localization of approximately doubly occupied Ni $e_g$  states makes it possible to split off the conduction band continuum to move closer to the valence band edge (see Fig.~\ref{fig:DOS_sketch}c), which is predominantly O $2p$, by an amount equivalent to the Coulomb repulsion energy $U$, thereby putting the Fermi-level just above the band-edge. This suggests that electrons donated by the In dopant are trapped locally on nearby purely Ni-sites, and do not contribute to conduction, making n-doped NiO a poor conductor. The inability of the $p$-$d$ hybridization to pin the otherwise empty Ni $e_g$ states close to the conduction band edge and the formation of deep donor levels as large as the Hubbard $U$ is a clear signature that ``local" electron addition and removal energies are dominated by the Coulomb $U$ interaction of the $d-d$  orbitals, i.e. the ``Mottness"\cite{phillips2010} of the material, even though the global bandgap is predominantly a $p-d$ type charge-transfer gap for excitations into the continuum. NiO therefore exhibits a special scenario of local Mott physics, even though in the continuum the spectra is dominated by a charge-transfer type $p-d$  gap. We note that recently it has been observed that incoherent charge-transfer excitations in Bi-based cuprates mimic localized Mott excitations at low doping levels~\cite{peli2017}.

\begin{acknowledgments}

F.W, C.S., H.S, H.N.L., P.G., J.T.K., A.Be., P.R.C.K., O.H., and A.Bh. were supported by the U.S. Department of Energy, Office of Science, Basic Energy Sciences, Materials Sciences and Engineering Division, as part of the Computational Materials Sciences Program and Center for Predictive Simulation of Functional Materials.
H.P. was supported by the U S Dept. of Energy, Office of Science, Basic Energy Sciences, Materials Sciences and Engineering Division.  
C.S. was supported in part (optical analysis) by Creative Materials Discovery Program through the National Research Foundation of Korea(NRF) funded by Ministry of Science and ICT(NRF-2017M3D1A1040828).
An award of computer time was provided by the Innovative and Novel Computational Impact on Theory and Experiment (INCITE) program. This research used resources of the Argonne Leadership Computing Facility, which is a DOE Office of Science User Facility supported under contract DE-AC02-06CH11357.
We gratefully acknowledge the computing resources provided on Bebop, a high-performance computing cluster operated by the Laboratory Computing Resource Center at Argonne National Laboratory.
Use of the Center for Nanoscale Materials, an Office of Science user facility, was supported by the U.S. Department of Energy, Office of Science, Office of Basic Energy Sciences, under Contract No. DE-AC02-06CH11357.
\end{acknowledgments}


\bibliographystyle{apsrev4-2}
\bibliography{KInNO}

\providecommand{\noopsort}[1]{}\providecommand{\singleletter}[1]{#1}%
\begin{thebibliography}{45}%
\makeatletter
\providecommand \@ifxundefined [1]{%
 \@ifx{#1\undefined}
}%
\providecommand \@ifnum [1]{%
 \ifnum #1\expandafter \@firstoftwo
 \else \expandafter \@secondoftwo
 \fi
}%
\providecommand \@ifx [1]{%
 \ifx #1\expandafter \@firstoftwo
 \else \expandafter \@secondoftwo
 \fi
}%
\providecommand \natexlab [1]{#1}%
\providecommand \enquote  [1]{``#1''}%
\providecommand \bibnamefont  [1]{#1}%
\providecommand \bibfnamefont [1]{#1}%
\providecommand \citenamefont [1]{#1}%
\providecommand \href@noop [0]{\@secondoftwo}%
\providecommand \href [0]{\begingroup \@sanitize@url \@href}%
\providecommand \@href[1]{\@@startlink{#1}\@@href}%
\providecommand \@@href[1]{\endgroup#1\@@endlink}%
\providecommand \@sanitize@url [0]{\catcode `\\12\catcode `\$12\catcode
  `\&12\catcode `\#12\catcode `\^12\catcode `\_12\catcode `\%12\relax}%
\providecommand \@@startlink[1]{}%
\providecommand \@@endlink[0]{}%
\providecommand \url  [0]{\begingroup\@sanitize@url \@url }%
\providecommand \@url [1]{\endgroup\@href {#1}{\urlprefix }}%
\providecommand \urlprefix  [0]{URL }%
\providecommand \Eprint [0]{\href }%
\providecommand \doibase [0]{https://doi.org/}%
\providecommand \selectlanguage [0]{\@gobble}%
\providecommand \bibinfo  [0]{\@secondoftwo}%
\providecommand \bibfield  [0]{\@secondoftwo}%
\providecommand \translation [1]{[#1]}%
\providecommand \BibitemOpen [0]{}%
\providecommand \bibitemStop [0]{}%
\providecommand \bibitemNoStop [0]{.\EOS\space}%
\providecommand \EOS [0]{\spacefactor3000\relax}%
\providecommand \BibitemShut  [1]{\csname bibitem#1\endcsname}%
\let\auto@bib@innerbib\@empty
\bibitem [{\citenamefont {{J. H. de Boer }}\ and\ \citenamefont
  {Verwey}(1937)}]{de1937}%
  \BibitemOpen
  \bibfield  {author} {\bibinfo {author} {\bibnamefont {{J. H. de Boer }}}and\
  \bibinfo {author} {\bibfnamefont {E.~J.~W.}\ \bibnamefont {Verwey}},\
  }\href@noop {} {\bibfield  {journal} {\bibinfo  {journal} {Proceedings of the
  Physical Society}\ }\textbf {\bibinfo {volume} {49}},\ \bibinfo {pages} {59}
  (\bibinfo {year} {1937})}\BibitemShut {NoStop}%
\bibitem [{\citenamefont {{N. F. Mott }}\ and\ \citenamefont
  {Peierls}(1937)}]{mott1937}%
  \BibitemOpen
  \bibfield  {author} {\bibinfo {author} {\bibnamefont {{N. F. Mott }}}and\
  \bibinfo {author} {\bibfnamefont {R.}~\bibnamefont {Peierls}},\ }\href
  {http://stacks.iop.org/0959-5309/49/i=4S/a=308} {\bibfield  {journal}
  {\bibinfo  {journal} {Proceedings of the Physical Society}\ }\textbf
  {\bibinfo {volume} {49}},\ \bibinfo {pages} {72} (\bibinfo {year}
  {1937})}\BibitemShut {NoStop}%
\bibitem [{\citenamefont {Lee}\ \emph {et~al.}(2006)\citenamefont {Lee},
  \citenamefont {Nagaosa},\ and\ \citenamefont {Wen}}]{lee2006}%
  \BibitemOpen
  \bibfield  {author} {\bibinfo {author} {\bibfnamefont {P.~A.}\ \bibnamefont
  {Lee}}, \bibinfo {author} {\bibfnamefont {N.}~\bibnamefont {Nagaosa}}, and\
  \bibinfo {author} {\bibfnamefont {X.-G.}\ \bibnamefont {Wen}},\ }\href@noop
  {} {\bibfield  {journal} {\bibinfo  {journal} {Reviews of Modern Physics}\
  }\textbf {\bibinfo {volume} {78}},\ \bibinfo {pages} {17} (\bibinfo {year}
  {2006})}\BibitemShut {NoStop}%
\bibitem [{\citenamefont {{I. H. Inoue }}\ and\ \citenamefont
  {Rozenberg}(2008)}]{inoue2008}%
  \BibitemOpen
  \bibfield  {author} {\bibinfo {author} {\bibnamefont {{I. H. Inoue }}}and\
  \bibinfo {author} {\bibfnamefont {M.~J.}\ \bibnamefont {Rozenberg}},\
  }\href@noop {} {\bibfield  {journal} {\bibinfo  {journal} {Advanced
  Functional Materials}\ }\textbf {\bibinfo {volume} {18}},\ \bibinfo {pages}
  {2289} (\bibinfo {year} {2008})}\BibitemShut {NoStop}%
\bibitem [{\citenamefont {Anisimov}\ \emph {et~al.}(1991)\citenamefont
  {Anisimov}, \citenamefont {Zaanen},\ and\ \citenamefont
  {Andersen}}]{anisimov1991}%
  \BibitemOpen
  \bibfield  {author} {\bibinfo {author} {\bibfnamefont {V.}~\bibnamefont
  {Anisimov}}, \bibinfo {author} {\bibfnamefont {J.}~\bibnamefont {Zaanen}},
  and\ \bibinfo {author} {\bibfnamefont {O.}~\bibnamefont {Andersen}},\
  }\href@noop {} {\bibfield  {journal} {\bibinfo  {journal} {Physical Review
  B}\ }\textbf {\bibinfo {volume} {44}},\ \bibinfo {pages} {943} (\bibinfo
  {year} {1991})}\BibitemShut {NoStop}%
\bibitem [{\citenamefont {{R. Newman }}\ and\ \citenamefont
  {Chrenko}(1959)}]{newman1959}%
  \BibitemOpen
  \bibfield  {author} {\bibinfo {author} {\bibnamefont {{R. Newman }}}and\
  \bibinfo {author} {\bibfnamefont {R.~M.}\ \bibnamefont {Chrenko}},\
  }\href@noop {} {\bibfield  {journal} {\bibinfo  {journal} {Physical Review}\
  }\textbf {\bibinfo {volume} {114}},\ \bibinfo {pages} {1507} (\bibinfo {year}
  {1959})}\BibitemShut {NoStop}%
\bibitem [{\citenamefont {{R. J. Powell }}\ and\ \citenamefont
  {Spicer}(1970)}]{powell1970}%
  \BibitemOpen
  \bibfield  {author} {\bibinfo {author} {\bibnamefont {{R. J. Powell }}}and\
  \bibinfo {author} {\bibfnamefont {W.~E.}\ \bibnamefont {Spicer}},\
  }\href@noop {} {\bibfield  {journal} {\bibinfo  {journal} {Physical Review
  B}\ }\textbf {\bibinfo {volume} {2}},\ \bibinfo {pages} {2182} (\bibinfo
  {year} {1970})}\BibitemShut {NoStop}%
\bibitem [{\citenamefont {Sasi}\ \emph {et~al.}(2002)\citenamefont {Sasi},
  \citenamefont {Gopchandran}, \citenamefont {Manoj}, \citenamefont {Koshy},
  \citenamefont {Rao},\ and\ \citenamefont {Vaidyan}}]{sasi2002}%
  \BibitemOpen
  \bibfield  {author} {\bibinfo {author} {\bibfnamefont {B.}~\bibnamefont
  {Sasi}}, \bibinfo {author} {\bibfnamefont {K.}~\bibnamefont {Gopchandran}},
  \bibinfo {author} {\bibfnamefont {P.}~\bibnamefont {Manoj}}, \bibinfo
  {author} {\bibfnamefont {P.}~\bibnamefont {Koshy}}, \bibinfo {author}
  {\bibfnamefont {P.~P.}\ \bibnamefont {Rao}}, and\ \bibinfo {author}
  {\bibfnamefont {V.}~\bibnamefont {Vaidyan}},\ }\href@noop {} {\bibfield
  {journal} {\bibinfo  {journal} {Vacuum}\ }\textbf {\bibinfo {volume} {68}},\
  \bibinfo {pages} {149} (\bibinfo {year} {2002})}\BibitemShut {NoStop}%
\bibitem [{\citenamefont {Zhang}\ \emph {et~al.}(2006)\citenamefont {Zhang},
  \citenamefont {Zhao},\ and\ \citenamefont {Zhu}}]{zhang2006}%
  \BibitemOpen
  \bibfield  {author} {\bibinfo {author} {\bibfnamefont {Z.}~\bibnamefont
  {Zhang}}, \bibinfo {author} {\bibfnamefont {Y.}~\bibnamefont {Zhao}}, and\
  \bibinfo {author} {\bibfnamefont {M.}~\bibnamefont {Zhu}},\ }\href@noop {}
  {\bibfield  {journal} {\bibinfo  {journal} {Applied Physics Letters}\
  }\textbf {\bibinfo {volume} {88}},\ \bibinfo {pages} {033101} (\bibinfo
  {year} {2006})}\BibitemShut {NoStop}%
\bibitem [{\citenamefont {Zaanen}\ \emph {et~al.}(1985)\citenamefont {Zaanen},
  \citenamefont {Sawatzky},\ and\ \citenamefont {Allen}}]{zaanen1985}%
  \BibitemOpen
  \bibfield  {author} {\bibinfo {author} {\bibfnamefont {J.}~\bibnamefont
  {Zaanen}}, \bibinfo {author} {\bibfnamefont {G.~A.}\ \bibnamefont
  {Sawatzky}}, and\ \bibinfo {author} {\bibfnamefont {J.~W.}\ \bibnamefont
  {Allen}},\ }\href {https://doi.org/10.1103/PhysRevLett.55.418} {\bibfield
  {journal} {\bibinfo  {journal} {Phys. Rev. Lett.}\ }\textbf {\bibinfo
  {volume} {55}},\ \bibinfo {pages} {418} (\bibinfo {year} {1985})}\BibitemShut
  {NoStop}%
\bibitem [{\citenamefont {Reinert}\ \emph {et~al.}(1995)\citenamefont
  {Reinert}, \citenamefont {Steiner}, \citenamefont {H{\"u}fner}, \citenamefont
  {Schmitt}, \citenamefont {Fink}, \citenamefont {Knupfer}, \citenamefont
  {Sandl},\ and\ \citenamefont {Bertel}}]{reinert1995}%
  \BibitemOpen
  \bibfield  {author} {\bibinfo {author} {\bibfnamefont {F.}~\bibnamefont
  {Reinert}}, \bibinfo {author} {\bibfnamefont {P.}~\bibnamefont {Steiner}},
  \bibinfo {author} {\bibfnamefont {S.}~\bibnamefont {H{\"u}fner}}, \bibinfo
  {author} {\bibfnamefont {H.}~\bibnamefont {Schmitt}}, \bibinfo {author}
  {\bibfnamefont {J.}~\bibnamefont {Fink}}, \bibinfo {author} {\bibfnamefont
  {M.}~\bibnamefont {Knupfer}}, \bibinfo {author} {\bibfnamefont
  {P.}~\bibnamefont {Sandl}}, and\ \bibinfo {author} {\bibfnamefont
  {E.}~\bibnamefont {Bertel}},\ }\href@noop {} {\bibfield  {journal} {\bibinfo
  {journal} {Zeitschrift f{\"u}r Physik B Condensed Matter}\ }\textbf {\bibinfo
  {volume} {97}},\ \bibinfo {pages} {83} (\bibinfo {year} {1995})}\BibitemShut
  {NoStop}%
\bibitem [{\citenamefont {Oka}\ \emph {et~al.}(2012)\citenamefont {Oka},
  \citenamefont {Yanagida}, \citenamefont {Nagashima}, \citenamefont {Kanai},
  \citenamefont {Xu}, \citenamefont {Park}, \citenamefont {Katayama-Yoshida},\
  and\ \citenamefont {Kawai}}]{oka2012}%
  \BibitemOpen
  \bibfield  {author} {\bibinfo {author} {\bibfnamefont {K.}~\bibnamefont
  {Oka}}, \bibinfo {author} {\bibfnamefont {T.}~\bibnamefont {Yanagida}},
  \bibinfo {author} {\bibfnamefont {K.}~\bibnamefont {Nagashima}}, \bibinfo
  {author} {\bibfnamefont {M.}~\bibnamefont {Kanai}}, \bibinfo {author}
  {\bibfnamefont {B.}~\bibnamefont {Xu}}, \bibinfo {author} {\bibfnamefont
  {B.~H.}\ \bibnamefont {Park}}, \bibinfo {author} {\bibfnamefont
  {H.}~\bibnamefont {Katayama-Yoshida}}, and\ \bibinfo {author} {\bibfnamefont
  {T.}~\bibnamefont {Kawai}},\ }\href@noop {} {\bibfield  {journal} {\bibinfo
  {journal} {Journal of the American Chemical Society}\ }\textbf {\bibinfo
  {volume} {134}},\ \bibinfo {pages} {2535} (\bibinfo {year}
  {2012})}\BibitemShut {NoStop}%
\bibitem [{\citenamefont {Kerli}\ \emph {et~al.}(2014)\citenamefont {Kerli},
  \citenamefont {Alver},\ and\ \citenamefont
  {Yayka{\c{s}}l{\i}}}]{kerli2014investigation}%
  \BibitemOpen
  \bibfield  {author} {\bibinfo {author} {\bibfnamefont {S.}~\bibnamefont
  {Kerli}}, \bibinfo {author} {\bibfnamefont {U.}~\bibnamefont {Alver}}, and\
  \bibinfo {author} {\bibfnamefont {H.}~\bibnamefont {Yayka{\c{s}}l{\i}}},\
  }\href@noop {} {\bibfield  {journal} {\bibinfo  {journal} {Applied Surface
  Science}\ }\textbf {\bibinfo {volume} {318}},\ \bibinfo {pages} {164}
  (\bibinfo {year} {2014})}\BibitemShut {NoStop}%
\bibitem [{\citenamefont {{Feinleib }}\ and\ \citenamefont
  {Adler}(1968)}]{feinleib1968}%
  \BibitemOpen
  \bibfield  {author} {\bibinfo {author} {\bibfnamefont {J.}~\bibnamefont
  {{Feinleib }}}and\ \bibinfo {author} {\bibfnamefont {D.}~\bibnamefont
  {Adler}},\ }\href@noop {} {\bibfield  {journal} {\bibinfo  {journal}
  {Physical Review Letters}\ }\textbf {\bibinfo {volume} {21}},\ \bibinfo
  {pages} {1010} (\bibinfo {year} {1968})}\BibitemShut {NoStop}%
\bibitem [{\citenamefont {Kuiper}\ \emph {et~al.}(1989)\citenamefont {Kuiper},
  \citenamefont {Kruizinga}, \citenamefont {Ghijsen}, \citenamefont
  {Sawatzky},\ and\ \citenamefont {Verweij}}]{kuiper1989character}%
  \BibitemOpen
  \bibfield  {author} {\bibinfo {author} {\bibfnamefont {P.}~\bibnamefont
  {Kuiper}}, \bibinfo {author} {\bibfnamefont {G.}~\bibnamefont {Kruizinga}},
  \bibinfo {author} {\bibfnamefont {J.}~\bibnamefont {Ghijsen}}, \bibinfo
  {author} {\bibfnamefont {G.~A.}\ \bibnamefont {Sawatzky}}, and\ \bibinfo
  {author} {\bibfnamefont {H.}~\bibnamefont {Verweij}},\ }\href@noop {}
  {\bibfield  {journal} {\bibinfo  {journal} {Physical Review Letters}\
  }\textbf {\bibinfo {volume} {62}},\ \bibinfo {pages} {221} (\bibinfo {year}
  {1989})}\BibitemShut {NoStop}%
\bibitem [{\citenamefont {H{\"u}fner}\ \emph {et~al.}(1992)\citenamefont
  {H{\"u}fner}, \citenamefont {Steiner}, \citenamefont {Reinert}, \citenamefont
  {Schmitt},\ and\ \citenamefont {Sandl}}]{hufner1992acceptors}%
  \BibitemOpen
  \bibfield  {author} {\bibinfo {author} {\bibfnamefont {S.}~\bibnamefont
  {H{\"u}fner}}, \bibinfo {author} {\bibfnamefont {P.}~\bibnamefont {Steiner}},
  \bibinfo {author} {\bibfnamefont {F.}~\bibnamefont {Reinert}}, \bibinfo
  {author} {\bibfnamefont {H.}~\bibnamefont {Schmitt}}, and\ \bibinfo {author}
  {\bibfnamefont {P.}~\bibnamefont {Sandl}},\ }\href@noop {} {\bibfield
  {journal} {\bibinfo  {journal} {Zeitschrift f{\"u}r Physik B Condensed
  Matter}\ }\textbf {\bibinfo {volume} {88}},\ \bibinfo {pages} {247} (\bibinfo
  {year} {1992})}\BibitemShut {NoStop}%
\bibitem [{\citenamefont {Van~Elp}\ \emph {et~al.}(1992)\citenamefont
  {Van~Elp}, \citenamefont {Eskes}, \citenamefont {Kuiper},\ and\ \citenamefont
  {Sawatzky}}]{van1992}%
  \BibitemOpen
  \bibfield  {author} {\bibinfo {author} {\bibfnamefont {J.}~\bibnamefont
  {Van~Elp}}, \bibinfo {author} {\bibfnamefont {H.}~\bibnamefont {Eskes}},
  \bibinfo {author} {\bibfnamefont {P.}~\bibnamefont {Kuiper}}, and\ \bibinfo
  {author} {\bibfnamefont {G.}~\bibnamefont {Sawatzky}},\ }\href@noop {}
  {\bibfield  {journal} {\bibinfo  {journal} {Physical Review B}\ }\textbf
  {\bibinfo {volume} {45}},\ \bibinfo {pages} {1612} (\bibinfo {year}
  {1992})}\BibitemShut {NoStop}%
\bibitem [{\citenamefont {Kune{\v{s}}}\ \emph
  {et~al.}(2007{\natexlab{a}})\citenamefont {Kune{\v{s}}}, \citenamefont
  {Anisimov}, \citenamefont {Lukoyanov},\ and\ \citenamefont
  {Vollhardt}}]{kunevs2007prb}%
  \BibitemOpen
  \bibfield  {author} {\bibinfo {author} {\bibfnamefont {J.}~\bibnamefont
  {Kune{\v{s}}}}, \bibinfo {author} {\bibfnamefont {V.}~\bibnamefont
  {Anisimov}}, \bibinfo {author} {\bibfnamefont {A.}~\bibnamefont {Lukoyanov}},
  and\ \bibinfo {author} {\bibfnamefont {D.}~\bibnamefont {Vollhardt}},\
  }\href@noop {} {\bibfield  {journal} {\bibinfo  {journal} {Physical Review
  B}\ }\textbf {\bibinfo {volume} {75}},\ \bibinfo {pages} {165115} (\bibinfo
  {year} {2007}{\natexlab{a}})}\BibitemShut {NoStop}%
\bibitem [{\citenamefont {Kune{\v{s}}}\ \emph
  {et~al.}(2007{\natexlab{b}})\citenamefont {Kune{\v{s}}}, \citenamefont
  {Anisimov}, \citenamefont {Skornyakov}, \citenamefont {Lukoyanov},\ and\
  \citenamefont {Vollhardt}}]{kunevs2007prl}%
  \BibitemOpen
  \bibfield  {author} {\bibinfo {author} {\bibfnamefont {J.}~\bibnamefont
  {Kune{\v{s}}}}, \bibinfo {author} {\bibfnamefont {V.}~\bibnamefont
  {Anisimov}}, \bibinfo {author} {\bibfnamefont {S.}~\bibnamefont
  {Skornyakov}}, \bibinfo {author} {\bibfnamefont {A.}~\bibnamefont
  {Lukoyanov}}, and\ \bibinfo {author} {\bibfnamefont {D.}~\bibnamefont
  {Vollhardt}},\ }\href@noop {} {\bibfield  {journal} {\bibinfo  {journal}
  {Physical Review Letters}\ }\textbf {\bibinfo {volume} {99}},\ \bibinfo
  {pages} {156404} (\bibinfo {year} {2007}{\natexlab{b}})}\BibitemShut
  {NoStop}%
\bibitem [{\citenamefont {{H. Chen }}\ and\ \citenamefont
  {Harding}(2012)}]{chen2012}%
  \BibitemOpen
  \bibfield  {author} {\bibinfo {author} {\bibnamefont {{H. Chen }}}and\
  \bibinfo {author} {\bibfnamefont {J.~H.}\ \bibnamefont {Harding}},\
  }\href@noop {} {\bibfield  {journal} {\bibinfo  {journal} {Physical Review
  B}\ }\textbf {\bibinfo {volume} {85}},\ \bibinfo {pages} {115127} (\bibinfo
  {year} {2012})}\BibitemShut {NoStop}%
\bibitem [{\citenamefont {Heyd}\ \emph {et~al.}(2005)\citenamefont {Heyd},
  \citenamefont {Peralta}, \citenamefont {Scuseria},\ and\ \citenamefont
  {Martin}}]{heyd2005}%
  \BibitemOpen
  \bibfield  {author} {\bibinfo {author} {\bibfnamefont {J.}~\bibnamefont
  {Heyd}}, \bibinfo {author} {\bibfnamefont {J.~E.}\ \bibnamefont {Peralta}},
  \bibinfo {author} {\bibfnamefont {G.~E.}\ \bibnamefont {Scuseria}}, and\
  \bibinfo {author} {\bibfnamefont {R.~L.}\ \bibnamefont {Martin}},\
  }\href@noop {} {\bibfield  {journal} {\bibinfo  {journal} {The Journal of
  Chemical Physics}\ }\textbf {\bibinfo {volume} {123}},\ \bibinfo {pages}
  {174101} (\bibinfo {year} {2005})}\BibitemShut {NoStop}%
\bibitem [{\citenamefont {Shinohara}\ \emph {et~al.}(2015)\citenamefont
  {Shinohara}, \citenamefont {Sharma}, \citenamefont {Dewhurst}, \citenamefont
  {Shallcross}, \citenamefont {Lathiotakis},\ and\ \citenamefont
  {Gross}}]{shinohara2015}%
  \BibitemOpen
  \bibfield  {author} {\bibinfo {author} {\bibfnamefont {Y.}~\bibnamefont
  {Shinohara}}, \bibinfo {author} {\bibfnamefont {S.}~\bibnamefont {Sharma}},
  \bibinfo {author} {\bibfnamefont {J.}~\bibnamefont {Dewhurst}}, \bibinfo
  {author} {\bibfnamefont {S.}~\bibnamefont {Shallcross}}, \bibinfo {author}
  {\bibfnamefont {N.}~\bibnamefont {Lathiotakis}}, and\ \bibinfo {author}
  {\bibfnamefont {E.}~\bibnamefont {Gross}},\ }\href@noop {} {\bibfield
  {journal} {\bibinfo  {journal} {New Journal of Physics}\ }\textbf {\bibinfo
  {volume} {17}},\ \bibinfo {pages} {093038} (\bibinfo {year}
  {2015})}\BibitemShut {NoStop}%
\bibitem [{\citenamefont {Mayer}(1997)}]{mayer1997}%
  \BibitemOpen
  \bibfield  {author} {\bibinfo {author} {\bibfnamefont {M.}~\bibnamefont
  {Mayer}},\ }\href {https://home.mpcdf.mpg.de/~mam/Report\%20IPP\%209-113.pdf}
  {\emph {\bibinfo {title} {SIMNRA User's Guide}}}\ (\bibinfo  {publisher}
  {Max-Planck-Institut fuer Plasmaphysik},\ \bibinfo {address} {Garching,
  Germany},\ \bibinfo {year} {1997})\BibitemShut {NoStop}%
\bibitem [{\citenamefont {Giannozzi}\ \emph {et~al.}(2009)\citenamefont
  {Giannozzi}, \citenamefont {Baroni}, \citenamefont {Bonini}, \citenamefont
  {Calandra}, \citenamefont {Car}, \citenamefont {Cavazzoni}, \citenamefont
  {Ceresoli}, \citenamefont {Chiarotti}, \citenamefont {Cococcioni},
  \citenamefont {Dabo}, \citenamefont {Corso}, \citenamefont {de~Gironcoli},
  \citenamefont {Fabris}, \citenamefont {Fratesi}, \citenamefont {Gebauer},
  \citenamefont {Gerstmann}, \citenamefont {Gougoussis}, \citenamefont
  {Kokalj}, \citenamefont {Lazzeri}, \citenamefont {M-.Samos}, \citenamefont
  {Mazari}, \citenamefont {Mauri}, \citenamefont {Mazzarello}, \citenamefont
  {Paolini}, \citenamefont {Pasquarello}, \citenamefont {Paulatto},
  \citenamefont {Sbraccia}, \citenamefont {Scandolo}, \citenamefont
  {Sclauzero}, \citenamefont {Seitsonen}, \citenamefont {Smogunov},
  \citenamefont {Umari},\ and\ \citenamefont {Wentzcovitch}}]{QE}%
  \BibitemOpen
  \bibfield  {author} {\bibinfo {author} {\bibfnamefont {P.}~\bibnamefont
  {Giannozzi}}, \bibinfo {author} {\bibfnamefont {S.}~\bibnamefont {Baroni}},
  \bibinfo {author} {\bibfnamefont {N.}~\bibnamefont {Bonini}}, \bibinfo
  {author} {\bibfnamefont {M.}~\bibnamefont {Calandra}}, \bibinfo {author}
  {\bibfnamefont {R.}~\bibnamefont {Car}}, \bibinfo {author} {\bibfnamefont
  {C.}~\bibnamefont {Cavazzoni}}, \bibinfo {author} {\bibfnamefont
  {D.}~\bibnamefont {Ceresoli}}, \bibinfo {author} {\bibfnamefont {G.~L.}\
  \bibnamefont {Chiarotti}}, \bibinfo {author} {\bibfnamefont {M.}~\bibnamefont
  {Cococcioni}}, \bibinfo {author} {\bibfnamefont {I.}~\bibnamefont {Dabo}},
  \bibinfo {author} {\bibfnamefont {A.~D.}\ \bibnamefont {Corso}}, \bibinfo
  {author} {\bibfnamefont {S.}~\bibnamefont {de~Gironcoli}}, \bibinfo {author}
  {\bibfnamefont {S.}~\bibnamefont {Fabris}}, \bibinfo {author} {\bibfnamefont
  {G.}~\bibnamefont {Fratesi}}, \bibinfo {author} {\bibfnamefont
  {R.}~\bibnamefont {Gebauer}}, \bibinfo {author} {\bibfnamefont
  {U.}~\bibnamefont {Gerstmann}}, \bibinfo {author} {\bibfnamefont
  {C.}~\bibnamefont {Gougoussis}}, \bibinfo {author} {\bibfnamefont
  {A.}~\bibnamefont {Kokalj}}, \bibinfo {author} {\bibfnamefont
  {M.}~\bibnamefont {Lazzeri}}, \bibinfo {author} {\bibfnamefont
  {L.}~\bibnamefont {M-.Samos}}, \bibinfo {author} {\bibfnamefont
  {N.}~\bibnamefont {Mazari}}, \bibinfo {author} {\bibfnamefont
  {F.}~\bibnamefont {Mauri}}, \bibinfo {author} {\bibfnamefont
  {R.}~\bibnamefont {Mazzarello}}, \bibinfo {author} {\bibfnamefont
  {S.}~\bibnamefont {Paolini}}, \bibinfo {author} {\bibfnamefont
  {A.}~\bibnamefont {Pasquarello}}, \bibinfo {author} {\bibfnamefont
  {L.}~\bibnamefont {Paulatto}}, \bibinfo {author} {\bibfnamefont
  {C.}~\bibnamefont {Sbraccia}}, \bibinfo {author} {\bibfnamefont
  {S.}~\bibnamefont {Scandolo}}, \bibinfo {author} {\bibfnamefont
  {G.}~\bibnamefont {Sclauzero}}, \bibinfo {author} {\bibfnamefont {A.~P.}\
  \bibnamefont {Seitsonen}}, \bibinfo {author} {\bibfnamefont {A.}~\bibnamefont
  {Smogunov}}, \bibinfo {author} {\bibfnamefont {P.}~\bibnamefont {Umari}},
  and\ \bibinfo {author} {\bibfnamefont {R.~M.}\ \bibnamefont {Wentzcovitch}},\
  }\href@noop {} {\bibfield  {journal} {\bibinfo  {journal} {J.\ Phys.:\
  Condens.\ Matter}\ }\textbf {\bibinfo {volume} {21}},\ \bibinfo {pages}
  {395502} (\bibinfo {year} {2009})}\BibitemShut {NoStop}%
\bibitem [{\citenamefont {Perdew}\ \emph {et~al.}(1996)\citenamefont {Perdew},
  \citenamefont {Burke},\ and\ \citenamefont {Ernzerhof}}]{PBE}%
  \BibitemOpen
  \bibfield  {author} {\bibinfo {author} {\bibfnamefont {J.~P.}\ \bibnamefont
  {Perdew}}, \bibinfo {author} {\bibfnamefont {K.}~\bibnamefont {Burke}}, and\
  \bibinfo {author} {\bibfnamefont {M.}~\bibnamefont {Ernzerhof}},\ }\href
  {https://doi.org/10.1103/PhysRevLett.77.3865} {\bibfield  {journal} {\bibinfo
   {journal} {Phys. Rev. Lett.}\ }\textbf {\bibinfo {volume} {77}},\ \bibinfo
  {pages} {3865} (\bibinfo {year} {1996})}\BibitemShut {NoStop}%
\bibitem [{\citenamefont {Shin}\ \emph {et~al.}(2017)\citenamefont {Shin},
  \citenamefont {Luo}, \citenamefont {Ganesh}, \citenamefont {Balachandran},
  \citenamefont {Krogel}, \citenamefont {Kent}, \citenamefont {Benali},\ and\
  \citenamefont {Heinonen}}]{shin2017}%
  \BibitemOpen
  \bibfield  {author} {\bibinfo {author} {\bibfnamefont {H.}~\bibnamefont
  {Shin}}, \bibinfo {author} {\bibfnamefont {Y.}~\bibnamefont {Luo}}, \bibinfo
  {author} {\bibfnamefont {P.}~\bibnamefont {Ganesh}}, \bibinfo {author}
  {\bibfnamefont {J.}~\bibnamefont {Balachandran}}, \bibinfo {author}
  {\bibfnamefont {J.~T.}\ \bibnamefont {Krogel}}, \bibinfo {author}
  {\bibfnamefont {P.~R.~C.}\ \bibnamefont {Kent}}, \bibinfo {author}
  {\bibfnamefont {A.}~\bibnamefont {Benali}}, and\ \bibinfo {author}
  {\bibfnamefont {O.}~\bibnamefont {Heinonen}},\ }\href
  {https://doi.org/10.1103/PhysRevMaterials.1.073603} {\bibfield  {journal}
  {\bibinfo  {journal} {Phys. Rev. Materials}\ }\textbf {\bibinfo {volume}
  {1}},\ \bibinfo {pages} {073603} (\bibinfo {year} {2017})}\BibitemShut
  {NoStop}%
\bibitem [{\citenamefont {Wrobel}\ \emph {et~al.}(2019)\citenamefont {Wrobel},
  \citenamefont {Shin}, \citenamefont {Sterbinsky}, \citenamefont {Hsiao},
  \citenamefont {Zuo}, \citenamefont {Ganesh}, \citenamefont {Krogel},
  \citenamefont {Benali}, \citenamefont {Kent}, \citenamefont {Heinonen} \emph
  {et~al.}}]{wrobel2019}%
  \BibitemOpen
  \bibfield  {author} {\bibinfo {author} {\bibfnamefont {F.}~\bibnamefont
  {Wrobel}}, \bibinfo {author} {\bibfnamefont {H.}~\bibnamefont {Shin}},
  \bibinfo {author} {\bibfnamefont {G.~E.}\ \bibnamefont {Sterbinsky}},
  \bibinfo {author} {\bibfnamefont {H.-W.}\ \bibnamefont {Hsiao}}, \bibinfo
  {author} {\bibfnamefont {J.-M.}\ \bibnamefont {Zuo}}, \bibinfo {author}
  {\bibfnamefont {P.}~\bibnamefont {Ganesh}}, \bibinfo {author} {\bibfnamefont
  {J.~T.}\ \bibnamefont {Krogel}}, \bibinfo {author} {\bibfnamefont
  {A.}~\bibnamefont {Benali}}, \bibinfo {author} {\bibfnamefont {P.~R.}\
  \bibnamefont {Kent}}, \bibinfo {author} {\bibfnamefont {O.}~\bibnamefont
  {Heinonen}},  \emph {et~al.},\ }\href@noop {} {\bibfield  {journal} {\bibinfo
   {journal} {Phys. Rev. Materials}\ }\textbf {\bibinfo {volume} {3}},\
  \bibinfo {pages} {115003} (\bibinfo {year} {2019})}\BibitemShut {NoStop}%
\bibitem [{\citenamefont {Burkatzki}\ \emph {et~al.}(2007)\citenamefont
  {Burkatzki}, \citenamefont {Filippi},\ and\ \citenamefont
  {Dolg}}]{burkatzki07}%
  \BibitemOpen
  \bibfield  {author} {\bibinfo {author} {\bibfnamefont {M.}~\bibnamefont
  {Burkatzki}}, \bibinfo {author} {\bibfnamefont {C.}~\bibnamefont {Filippi}},
  and\ \bibinfo {author} {\bibfnamefont {M.}~\bibnamefont {Dolg}},\ }\href@noop
  {} {\bibfield  {journal} {\bibinfo  {journal} {J.\ Chem.\ Phys.}\ }\textbf
  {\bibinfo {volume} {126}},\ \bibinfo {pages} {234105} (\bibinfo {year}
  {2007})}\BibitemShut {NoStop}%
\bibitem [{\citenamefont {Burkatzki}\ \emph {et~al.}(2008)\citenamefont
  {Burkatzki}, \citenamefont {Filippi},\ and\ \citenamefont
  {Dolg}}]{burkatzki08}%
  \BibitemOpen
  \bibfield  {author} {\bibinfo {author} {\bibfnamefont {M.}~\bibnamefont
  {Burkatzki}}, \bibinfo {author} {\bibfnamefont {C.}~\bibnamefont {Filippi}},
  and\ \bibinfo {author} {\bibfnamefont {M.}~\bibnamefont {Dolg}},\ }\href@noop
  {} {\bibfield  {journal} {\bibinfo  {journal} {J.\ Chem.\ Phys.}\ }\textbf
  {\bibinfo {volume} {129}},\ \bibinfo {pages} {164115} (\bibinfo {year}
  {2008})}\BibitemShut {NoStop}%
\bibitem [{\citenamefont {Haule}\ \emph {et~al.}(2010)\citenamefont {Haule},
  \citenamefont {Yee},\ and\ \citenamefont {Kim}}]{PRB.81.195107}%
  \BibitemOpen
  \bibfield  {author} {\bibinfo {author} {\bibfnamefont {K.}~\bibnamefont
  {Haule}}, \bibinfo {author} {\bibfnamefont {C.-H.}\ \bibnamefont {Yee}}, and\
  \bibinfo {author} {\bibfnamefont {K.}~\bibnamefont {Kim}},\ }\href
  {https://doi.org/10.1103/PhysRevB.81.195107} {\bibfield  {journal} {\bibinfo
  {journal} {Phys. Rev. B}\ }\textbf {\bibinfo {volume} {81}},\ \bibinfo
  {pages} {195107} (\bibinfo {year} {2010})}\BibitemShut {NoStop}%
\bibitem [{\citenamefont {Haule}(2007)}]{PRB.75.155113}%
  \BibitemOpen
  \bibfield  {author} {\bibinfo {author} {\bibfnamefont {K.}~\bibnamefont
  {Haule}},\ }\href {https://doi.org/10.1103/PhysRevB.75.155113} {\bibfield
  {journal} {\bibinfo  {journal} {Phys. Rev. B}\ }\textbf {\bibinfo {volume}
  {75}},\ \bibinfo {pages} {155113} (\bibinfo {year} {2007})}\BibitemShut
  {NoStop}%
\bibitem [{\citenamefont {Gull}\ \emph {et~al.}(2011)\citenamefont {Gull},
  \citenamefont {Millis}, \citenamefont {Lichtenstein}, \citenamefont
  {Rubtsov}, \citenamefont {Troyer},\ and\ \citenamefont
  {Werner}}]{RMP.83.349}%
  \BibitemOpen
  \bibfield  {author} {\bibinfo {author} {\bibfnamefont {E.}~\bibnamefont
  {Gull}}, \bibinfo {author} {\bibfnamefont {A.~J.}\ \bibnamefont {Millis}},
  \bibinfo {author} {\bibfnamefont {A.~I.}\ \bibnamefont {Lichtenstein}},
  \bibinfo {author} {\bibfnamefont {A.~N.}\ \bibnamefont {Rubtsov}}, \bibinfo
  {author} {\bibfnamefont {M.}~\bibnamefont {Troyer}}, and\ \bibinfo {author}
  {\bibfnamefont {P.}~\bibnamefont {Werner}},\ }\href
  {https://doi.org/10.1103/RevModPhys.83.349} {\bibfield  {journal} {\bibinfo
  {journal} {Rev. Mod. Phys.}\ }\textbf {\bibinfo {volume} {83}},\ \bibinfo
  {pages} {349} (\bibinfo {year} {2011})}\BibitemShut {NoStop}%
\bibitem [{\citenamefont {Haule}\ \emph {et~al.}(2014)\citenamefont {Haule},
  \citenamefont {Birol},\ and\ \citenamefont {Kotliar}}]{PRB.90.075136}%
  \BibitemOpen
  \bibfield  {author} {\bibinfo {author} {\bibfnamefont {K.}~\bibnamefont
  {Haule}}, \bibinfo {author} {\bibfnamefont {T.}~\bibnamefont {Birol}}, and\
  \bibinfo {author} {\bibfnamefont {G.}~\bibnamefont {Kotliar}},\ }\href
  {https://doi.org/10.1103/PhysRevB.90.075136} {\bibfield  {journal} {\bibinfo
  {journal} {Phys. Rev. B}\ }\textbf {\bibinfo {volume} {90}},\ \bibinfo
  {pages} {075136} (\bibinfo {year} {2014})}\BibitemShut {NoStop}%
\bibitem [{\citenamefont {Haule}(2015)}]{PRL.115.196403}%
  \BibitemOpen
  \bibfield  {author} {\bibinfo {author} {\bibfnamefont {K.}~\bibnamefont
  {Haule}},\ }\href {https://doi.org/10.1103/PhysRevLett.115.196403} {\bibfield
   {journal} {\bibinfo  {journal} {Phys. Rev. Lett.}\ }\textbf {\bibinfo
  {volume} {115}},\ \bibinfo {pages} {196403} (\bibinfo {year}
  {2015})}\BibitemShut {NoStop}%
\bibitem [{\citenamefont {Koyama}\ \emph {et~al.}(2005)\citenamefont {Koyama},
  \citenamefont {Mizoguchi}, \citenamefont {Ikeno},\ and\ \citenamefont
  {Tanaka}}]{koyama2005}%
  \BibitemOpen
  \bibfield  {author} {\bibinfo {author} {\bibfnamefont {Y.}~\bibnamefont
  {Koyama}}, \bibinfo {author} {\bibfnamefont {T.}~\bibnamefont {Mizoguchi}},
  \bibinfo {author} {\bibfnamefont {H.}~\bibnamefont {Ikeno}}, and\ \bibinfo
  {author} {\bibfnamefont {I.}~\bibnamefont {Tanaka}},\ }\href
  {https://doi.org/10.1021/jp050486b} {\bibfield  {journal} {\bibinfo
  {journal} {The Journal of Physical Chemistry B}\ }\textbf {\bibinfo {volume}
  {109}},\ \bibinfo {pages} {10749} (\bibinfo {year} {2005})},\ \bibinfo {note}
  {pMID: 16852306},\ \Eprint
  {https://arxiv.org/abs/https://doi.org/10.1021/jp050486b}
  {https://doi.org/10.1021/jp050486b} \BibitemShut {NoStop}%
\bibitem [{\citenamefont {Wooton}(1972)}]{wooton1972}%
  \BibitemOpen
  \bibfield  {author} {\bibinfo {author} {\bibfnamefont {F.}~\bibnamefont
  {Wooton}},\ }\href@noop {} {\emph {\bibinfo {title} {Optical Properties of
  Solids}}}\ (\bibinfo  {publisher} {Academic Press},\ \bibinfo {address} {111
  Fifth Avenue, New York, New York 10003},\ \bibinfo {year} {1972})\BibitemShut
  {NoStop}%
\bibitem [{\citenamefont {Jellison}\ and\ \citenamefont
  {Modine}(1996)}]{jellison1996}%
  \BibitemOpen
  \bibfield  {author} {\bibinfo {author} {\bibfnamefont {G.~E.}\ \bibnamefont
  {Jellison}}and\ \bibinfo {author} {\bibfnamefont {F.~A.}\ \bibnamefont
  {Modine}},\ }\href {https://doi.org/10.1063/1.118064} {\bibfield  {journal}
  {\bibinfo  {journal} {Applied Physics Letters}\ }\textbf {\bibinfo {volume}
  {69}},\ \bibinfo {pages} {371} (\bibinfo {year} {1996})}\BibitemShut
  {NoStop}%
\bibitem [{\citenamefont {Herman}(1996)}]{herman1996}%
  \BibitemOpen
  \bibfield  {author} {\bibinfo {author} {\bibfnamefont {I.~P.}\ \bibnamefont
  {Herman}},\ }\href@noop {} {\emph {\bibinfo {title} {Optical Diagnostics for
  Thin Film Processing}}}\ (\bibinfo  {publisher} {Academic Press},\ \bibinfo
  {address} {525 B Street, Suite 1900, San Diego, California 92101-4495, USA},\
  \bibinfo {year} {1996})\BibitemShut {NoStop}%
\bibitem [{\citenamefont {Tauc}\ \emph {et~al.}(1966)\citenamefont {Tauc},
  \citenamefont {Grigorovici},\ and\ \citenamefont {Vancu}}]{tauc1966}%
  \BibitemOpen
  \bibfield  {author} {\bibinfo {author} {\bibfnamefont {J.}~\bibnamefont
  {Tauc}}, \bibinfo {author} {\bibfnamefont {R.}~\bibnamefont {Grigorovici}},
  and\ \bibinfo {author} {\bibfnamefont {A.}~\bibnamefont {Vancu}},\ }\href
  {https://doi.org/10.1002/pssb.19660150224} {\bibfield  {journal} {\bibinfo
  {journal} {Physica Status Solidi (b)}\ }\textbf {\bibinfo {volume} {15}},\
  \bibinfo {pages} {627} (\bibinfo {year} {1966})}\BibitemShut {NoStop}%
\bibitem [{\citenamefont {{H. G. Tompkins }}\ and\ \citenamefont
  {Irene}(2005)}]{tompkins2005}%
  \BibitemOpen
  \bibfield  {author} {\bibinfo {author} {\bibnamefont {{H. G. Tompkins }}}and\
  \bibinfo {author} {\bibfnamefont {E.~A.}\ \bibnamefont {Irene}},\ }\href@noop
  {} {\emph {\bibinfo {title} {Handbook of Ellipsometry}}}\ (\bibinfo
  {publisher} {William Andrew publishing},\ \bibinfo {address} {13 Eaton
  Avenue, Norwich, New York 13815},\ \bibinfo {year} {2005})\BibitemShut
  {NoStop}%
\bibitem [{\citenamefont {Biermann}\ \emph {et~al.}(2005)\citenamefont
  {Biermann}, \citenamefont {Poteryaev}, \citenamefont {Lichtenstein},\ and\
  \citenamefont {Georges}}]{biermann2005}%
  \BibitemOpen
  \bibfield  {author} {\bibinfo {author} {\bibfnamefont {S.}~\bibnamefont
  {Biermann}}, \bibinfo {author} {\bibfnamefont {A.}~\bibnamefont {Poteryaev}},
  \bibinfo {author} {\bibfnamefont {A.~I.}\ \bibnamefont {Lichtenstein}}, and\
  \bibinfo {author} {\bibfnamefont {A.}~\bibnamefont {Georges}},\ }\href
  {https://doi.org/10.1103/PhysRevLett.94.026404} {\bibfield  {journal}
  {\bibinfo  {journal} {Phys. Rev. Lett.}\ }\textbf {\bibinfo {volume} {94}},\
  \bibinfo {pages} {026404} (\bibinfo {year} {2005})}\BibitemShut {NoStop}%
\bibitem [{\citenamefont {Giovannetti}\ \emph {et~al.}(2014)\citenamefont
  {Giovannetti}, \citenamefont {Aichhorn},\ and\ \citenamefont
  {Capone}}]{giovannetti2014}%
  \BibitemOpen
  \bibfield  {author} {\bibinfo {author} {\bibfnamefont {G.}~\bibnamefont
  {Giovannetti}}, \bibinfo {author} {\bibfnamefont {M.}~\bibnamefont
  {Aichhorn}}, and\ \bibinfo {author} {\bibfnamefont {M.}~\bibnamefont
  {Capone}},\ }\href@noop {} {\bibfield  {journal} {\bibinfo  {journal}
  {Physical Review B}\ }\textbf {\bibinfo {volume} {90}},\ \bibinfo {pages}
  {245134} (\bibinfo {year} {2014})}\BibitemShut {NoStop}%
\bibitem [{\citenamefont {Vasudevan}\ \emph {et~al.}(2018)\citenamefont
  {Vasudevan}, \citenamefont {Dixit}, \citenamefont {Tselev}, \citenamefont
  {Qiao}, \citenamefont {Meyer}, \citenamefont {Cooper}, \citenamefont
  {Baddorf}, \citenamefont {Lee}, \citenamefont {Ganesh},\ and\ \citenamefont
  {Kalinin}}]{vasudevan2018}%
  \BibitemOpen
  \bibfield  {author} {\bibinfo {author} {\bibfnamefont {R.~K.}\ \bibnamefont
  {Vasudevan}}, \bibinfo {author} {\bibfnamefont {H.}~\bibnamefont {Dixit}},
  \bibinfo {author} {\bibfnamefont {A.}~\bibnamefont {Tselev}}, \bibinfo
  {author} {\bibfnamefont {L.}~\bibnamefont {Qiao}}, \bibinfo {author}
  {\bibfnamefont {T.~L.}\ \bibnamefont {Meyer}}, \bibinfo {author}
  {\bibfnamefont {V.~R.}\ \bibnamefont {Cooper}}, \bibinfo {author}
  {\bibfnamefont {A.~P.}\ \bibnamefont {Baddorf}}, \bibinfo {author}
  {\bibfnamefont {H.~N.}\ \bibnamefont {Lee}}, \bibinfo {author} {\bibfnamefont
  {P.}~\bibnamefont {Ganesh}}, and\ \bibinfo {author} {\bibfnamefont {S.~V.}\
  \bibnamefont {Kalinin}},\ }\href
  {https://doi.org/10.1103/PhysRevMaterials.2.104418} {\bibfield  {journal}
  {\bibinfo  {journal} {Phys. Rev. Materials}\ }\textbf {\bibinfo {volume}
  {2}},\ \bibinfo {pages} {104418} (\bibinfo {year} {2018})}\BibitemShut
  {NoStop}%
\bibitem [{\citenamefont {Phillips}(2010)}]{phillips2010}%
  \BibitemOpen
  \bibfield  {author} {\bibinfo {author} {\bibfnamefont {P.}~\bibnamefont
  {Phillips}},\ }\href@noop {} {\bibfield  {journal} {\bibinfo  {journal}
  {Reviews of Modern Physics}\ }\textbf {\bibinfo {volume} {82}},\ \bibinfo
  {pages} {1719} (\bibinfo {year} {2010})}\BibitemShut {NoStop}%
\bibitem [{\citenamefont {Peli}\ \emph {et~al.}(2017)\citenamefont {Peli},
  \citenamefont {Dal~Conte}, \citenamefont {Comin}, \citenamefont {Nembrini},
  \citenamefont {Ronchi}, \citenamefont {Abrami}, \citenamefont {Banfi},
  \citenamefont {Ferrini}, \citenamefont {Brida}, \citenamefont {Lupi} \emph
  {et~al.}}]{peli2017}%
  \BibitemOpen
  \bibfield  {author} {\bibinfo {author} {\bibfnamefont {S.}~\bibnamefont
  {Peli}}, \bibinfo {author} {\bibfnamefont {S.}~\bibnamefont {Dal~Conte}},
  \bibinfo {author} {\bibfnamefont {R.}~\bibnamefont {Comin}}, \bibinfo
  {author} {\bibfnamefont {N.}~\bibnamefont {Nembrini}}, \bibinfo {author}
  {\bibfnamefont {A.}~\bibnamefont {Ronchi}}, \bibinfo {author} {\bibfnamefont
  {P.}~\bibnamefont {Abrami}}, \bibinfo {author} {\bibfnamefont
  {F.}~\bibnamefont {Banfi}}, \bibinfo {author} {\bibfnamefont
  {G.}~\bibnamefont {Ferrini}}, \bibinfo {author} {\bibfnamefont
  {D.}~\bibnamefont {Brida}}, \bibinfo {author} {\bibfnamefont
  {S.}~\bibnamefont {Lupi}},  \emph {et~al.},\ }\href@noop {} {\bibfield
  {journal} {\bibinfo  {journal} {Nature Physics}\ }\textbf {\bibinfo {volume}
  {13}},\ \bibinfo {pages} {806} (\bibinfo {year} {2017})}\BibitemShut
  {NoStop}%
\end{thebibliography}%

\section*{Appendix}
\section*{Rutherford Backscattering}

\begin{figure}[h]
\includegraphics[width=8cm]{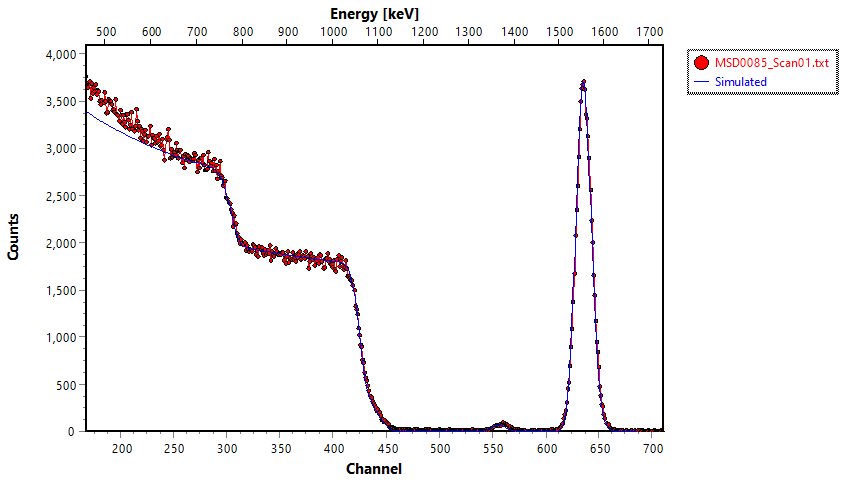}
\centering
\caption{\label{fig:RBSK} RBS spectrum (red dots) and fit to the data (blue line) of NiO doped with 3.6\,\% potassium.} 
\end{figure}

\begin{figure}[h]
\includegraphics[width=8cm]{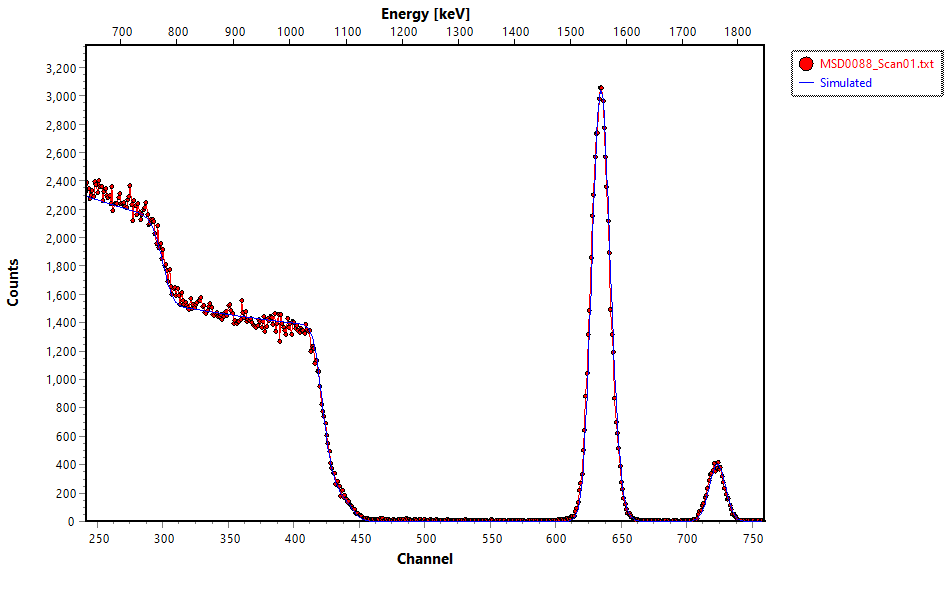}
\centering
\caption{\label{fig:RBSIn} RBS spectrum (red dots) and fit to the data (blue line) of NiO doped with 3.7\,\% indium.} 
\end{figure}

To determine the doping level of our samples, we measured RBS and analyzed the data with the SIMNRA software~\cite{mayer1997}, which yields the number of atoms per area. While the absolute number of atoms per area can have a relatively large error for samples with a low number of atoms, especially for the dopants potassium and indium in this case, the ratio of dopant to nickel, and thereby the doping level, is very accurate. We estimate an upper bound for the error to be $\pm$2 percentage points in terms of doping level.

Exemplary spectra for potassium and indium doped samples are shown in Fig.~\ref{fig:RBSK} and Fig.~\ref{fig:RBSIn}


\section*{Sample thicknesses}

We measured the sample thicknesses with x-ray reflectivity and the results are displayed in Table~\ref{tab:thickness}.

\begin{table}[]
\caption{\label{tab:thickness}
Thicknesses of the samples used in this study determined by x-ray reflectivity.}
\centering
\begin{tabular}{@{}cc@{}}
\hline
\hline
doping (\%) & thickness (nm) \\ \hline
9.1         & 22             \\
5.0         & 22             \\
3.6         & 23             \\
0           & 24             \\
-3.7        & 24             \\
-7.8        & 22             \\
-15.6       & 23             \\ 
\hline
\hline
\end{tabular}%
\end{table}

\FloatBarrier

\section*{Li-doped NiO}
\begin{figure}
    \centering
    \includegraphics[width=\columnwidth]{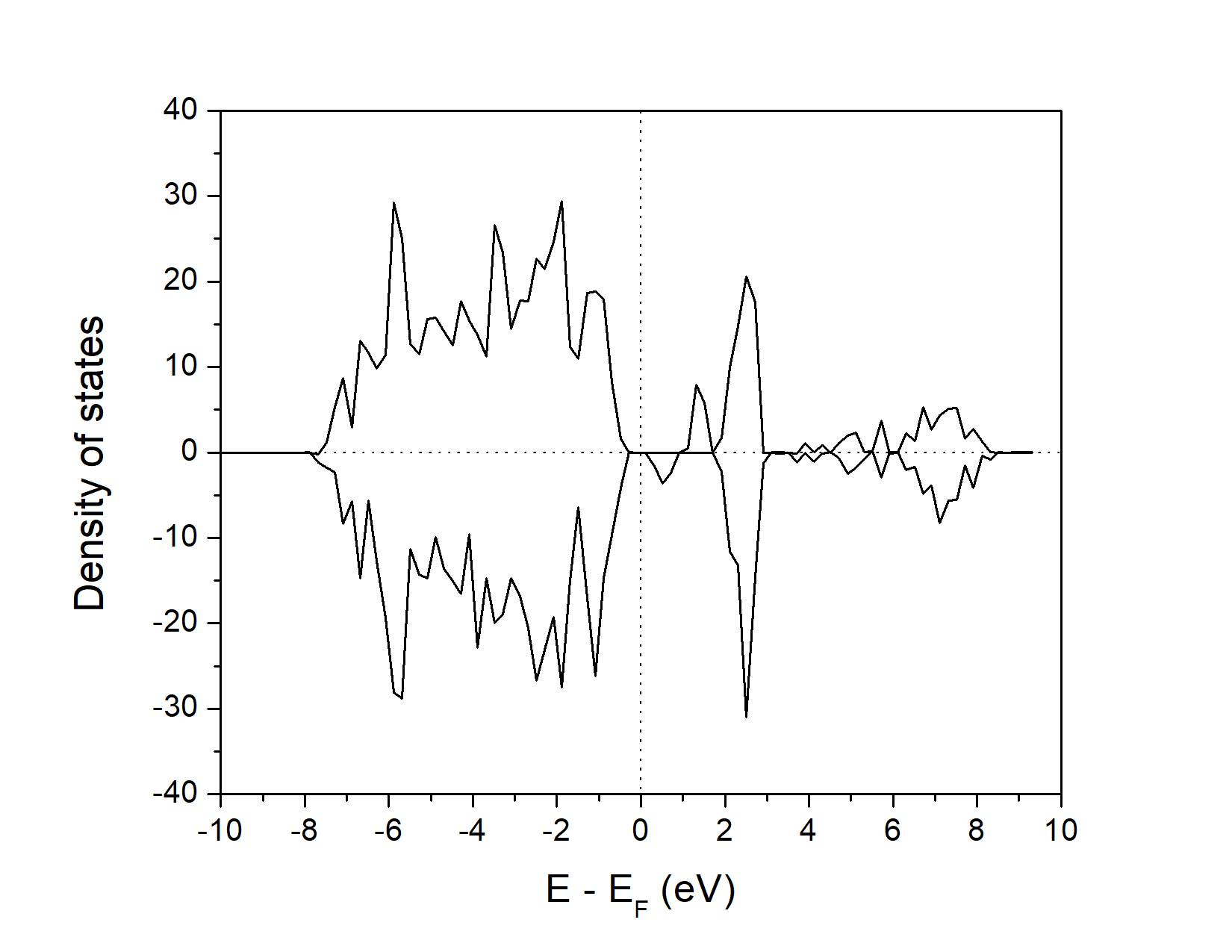}
    \caption{Calculated PBE+U density-of-states for a 32-atom NiO supercell with one substitutional Li dopant with up-spin DOS on top and down-spin DOS below.}
    \label{fig:Li-DOS}
\end{figure}

As described in Sec.\ref{subsec:DFT+DMFT} we performed PBE+U calculations on a 32-atom NiO supercell with one substitutional Li dopant, corresponding to about 6\% doping. The DOS for Li-doping is shown in Fig.\ref{fig:Li-DOS}. Defect states are above the valence band edge in the gap, in agreement with Chen and Harding\cite{chen2012}.

\section*{Substititional and interstitial In-doping}
We performed PBE+U calculations of In-defect creation energies using 32-atom NiO supercells. In one calculation, we substituted In for one Ni, in another calculation, we introduced an interstitial In defect. We relaxed the structures as described in Sec.~\ref{subsec:DFT+DMFT} and calculated defect creation energies. The result is that the interstitial defect creation energy is almost seven eV larger than the substitutional defect creation energy. Therefore, we conclude that In dopes substitutionally, certainly for the relatively low concentrations considered here.

\section*{Bond lengths}
The calculated dopant-oxygen bond lengths for the relaxed substitutionally doped structures (GGA+U) are as follows. 
\begin{itemize}
    \item K-doped NiO: near neighbor oxygen are symmetrically placed around the K dopant with a bond length of 2.355~{\AA}.
    \item In-doped NiO: the oxygen octahedron around the In dopant is slightly disorted with two In-O bonds 2.204~{\AA} and four In-O bonds 2.195~{\AA}. 
    \item Li-doped NiO: the oxygen octahedron is significantly distorted with two Li-O bonds 2.093~{\AA} and four bonds 2.288~{\AA}. 
\end{itemize}
\vspace{12pt}

\section*{L\"owdin charges}
L\"owdin charges were calculated using Quantum Espresso.  For the pure NiO and In-doped NiO, the calculations were using the same parameters and pseudopotentials as described in the main text. The number of valence electrons and atomic configurations were 10 ([Ar] 4s2 3d8 4p0), 6 ([He] 2s2 2p4), and 13 ([Kr] 5s2 5p1 4d10), for Ni, O, and In, respectively. For the K-doped L\"owdin charges, we used a USPP pseudopotential with nine valence electrons and a  [Ne] 3s2 3p6 4s1 configuration on the geometry relaxed as described in the main text. The L\"owdin charge analysis is shown below.

\begin{widetext}
\begin{figure}[h]
\includegraphics[width=25cm]{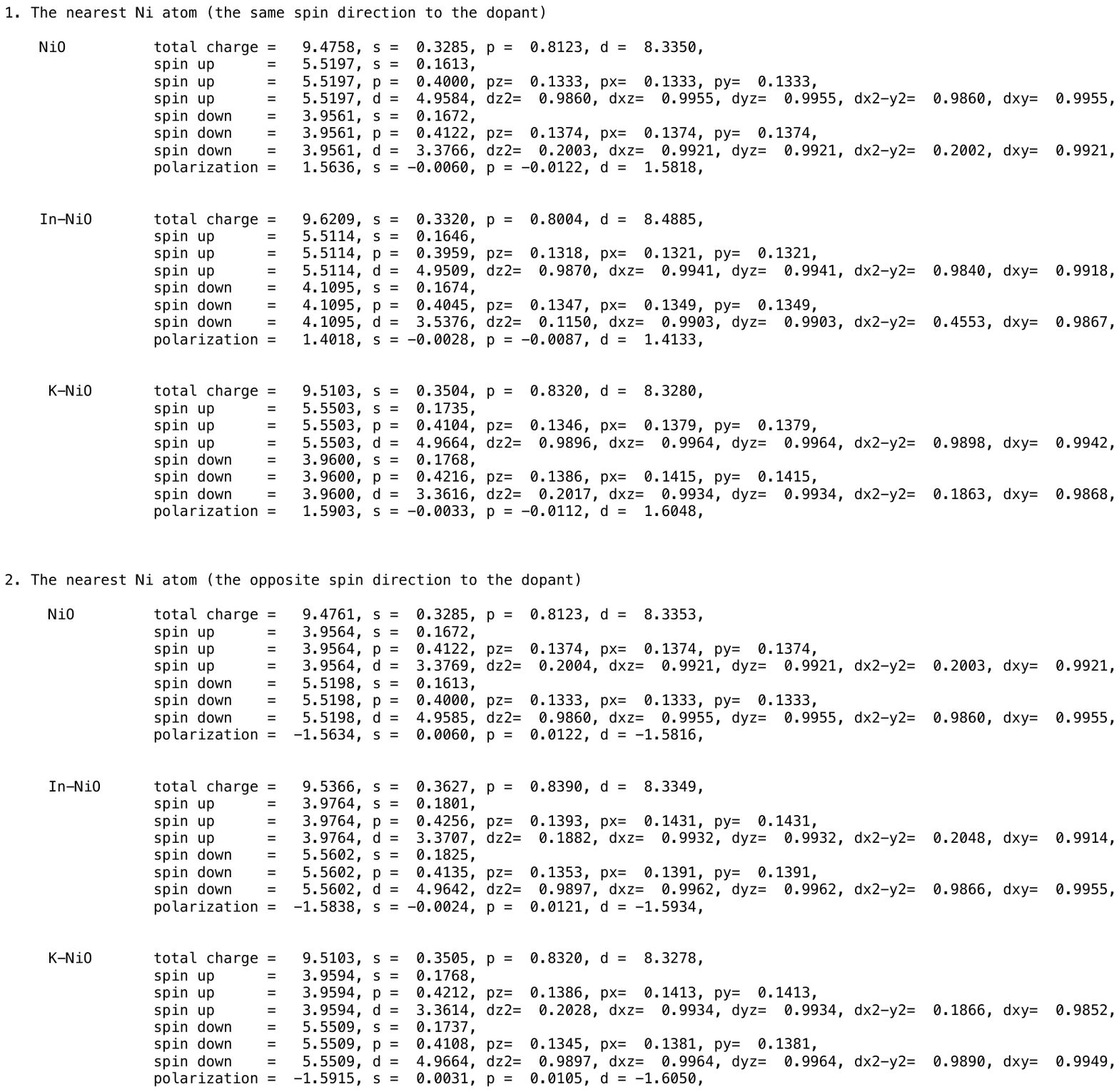}
\centering
\caption{\label{fig:Lowdin} L\"owdin charge analysis for NiO, K-doped NiO, and In-doped NiO.} 
\end{figure}
\end{widetext}

\end{document}